\documentclass[%
 reprint,
superscriptaddress,
 amsmath,amssymb,
 aps,
floatfix,y
]{revtex4-2}
\usepackage{graphicx}
\usepackage{dcolumn}
\usepackage{bm}
\usepackage{pgfplots}
\usepackage{siunitx}
\pgfplotsset{compat=1.17} \usepackage[breaklinks,colorlinks,bookmarks=false,citecolor=blue,linkcolor=red,urlcolor=blue]{hyperref}

\begin{document}


\title{
The percolating cluster is invisible to image recognition with deep learning
}

\author{Dj\'enabou Bayo}
\email{Djenabou.Bayo@warwick.ac.uk}
\affiliation{Department of Physics, University of Warwick, Coventry, CV4 7AL, United Kingdom}
\affiliation{Laboratoire de Physique Th\'eorique et Mod\'elisation, CNRS UMR 8089, CY Cergy Paris Universit\'e, Cergy-Pontoise, France}


\author{Andreas Honecker}
\email{Andreas.Honecker@cyu.fr}
\affiliation{Laboratoire de Physique Th\'eorique et Mod\'elisation, CNRS UMR 8089, CY Cergy Paris Universit\'e, Cergy-Pontoise, France}%

\author{Rudolf A. R\"omer}
\email{R.Roemer@warwick.ac.uk}
\affiliation{Department of Physics, University of Warwick, Coventry, CV4 7AL, United Kingdom}%


\date{\today}

\begin{abstract}
We study the two-dimensional site-percolation model on a square lattice. In this paradigmatic model, sites are randomly occupied with probability $p$; a second-order phase transition from a non-percolating to a fully percolating phase appears at occupation density $p_c$, called percolation threshold. 
Through supervised deep learning approaches like classification and regression, we show that standard convolutional neural networks (CNNs), known to work well in similar image recognition tasks, can identify $p_c$ and indeed classify the states of a percolation lattice according to their $p$ content or predict their $p$ value via regression. When using instead of $p$ the spatial cluster correlation length $\xi$ as labels, the recognition is beginning to falter. 
Finally, we show that the same network struggles to detect the presence of a spanning cluster. Rather, predictive power seems lost and the absence or presence of a \emph{global} spanning cluster is not noticed by a CNN with \emph{local} convolutional kernel. Since the existence of such a spanning cluster is at the heart of the percolation problem, our results suggest that CNNs require careful application when used in physics, particularly when encountering less-explored situations.

\end{abstract}

\maketitle


\section{\label{sec:introduction}Introduction}

Convolution neural nets (CNN) are a class of deep, i.e., multi-layered, neural nets (DNNs) in which spatial locality of data values is retained during training in a machine learning (ML) setting. When coupled with a form of residual learning \cite{He2016DeepRecognition}, the resulting residual networks ({\sc ResNets}) have been shown to allow astonishing precision when classifying images, e.g., of animals \cite{Tabak2019} and handwritten characters \cite{Zhang2017CombinationRecognition}, or when predicting numerical values, e.g., of market prices \cite{Zhao2020}. 
In recent years, we also witnessed the emergence of DNN techniques in several fields of physics as a new tool for data analysis \cite{Nomura2017a,Rao2018MachineSystems,Zhang2017a,Stoudenmire2017}. In condensed matter physics in particular, DNN and CNN proved to be performing well in identifying and classifying phases of material states \cite{Carrasquilla2017,doi:10.7566/JPSJ.89.022001,PhysRevX.7.031038,Venderley2018b}.

Despite all these studies, the ML process in itself tends to be somewhat a black box, and it is yet not known what is allowing a DNN to correctly identify a phase.
In order to gain further insight into this issue, we choose a well-known and well-studied system exhibiting perhaps the simplest of all second-order phase transitions, the \emph{site-percolation} model in two spatial dimensions \cite{Broadbent1957a,Stauffer1991IntroductionTheory}. In this model, a cluster spanning throughout the system emerges at an occupation probability $p_c$, leading to a non-spanning phase when $p < p_c$ while $p \geq p_c$ corresponds to the phase with at least one such spanning cluster \cite{Stauffer1991IntroductionTheory}.
Several ML studies on the percolation model have been already published, mostly using \emph{supervised} learning in order to identify the two phases via ML classification \cite{Zhang2019,Shen2021SupervisedPercolation}. An estimate of the critical exponent, $\nu$, of the percolation transition has also been given \cite{Zhang2019}. The task of determining $p_c$ was further used to evaluate different ML regression techniques in Ref.\ \cite{Patwardhan2022MachineNetworks}.
For \emph{unsupervised} and \emph{generative} learning, less work has been done \cite{Shen2021SupervisedPercolation,Yu2020UnsupervisedPercolation}. While some successes have been reported \cite{Yu2020UnsupervisedPercolation}, other works show the complexities involved when trying to predict percolation states \cite{Shen2021SupervisedPercolation}.

In this work, we start by replicating some of the supervised DL analyses. We find that CNNs usually employed in image recognition ML tasks also work very well for classifying percolation states according to $p$ as well as for regression when determining $p$ from such states. 
The results are less convincing when instead of $p$, we use the spatial correlation lengths $\xi$ as an alternative means to characterize the phase boundary. We find that, even when correcting for probable difficulties due to non-balanced data availability for $\xi$, classification and regression tasks fail to give acceptably diagonal confusion matrices. 
Crucially, when analyzing in detail whether spanning clusters $< p_c$ or non-spanning clusters $> p_c$ are correctly identified, we find the CNNs that performed so well in the initial image recognition tasks now consistently fail to reflect the ground truth. Rather, it appears that the CNNs use $p$ as a proxy measure to inform their classification predictions --- a strategy that is obviously false for the percolation problem. We confirm this conclusion by testing our networks with bespoke test sets that include artificially spanning clusters $< p_c$ or firebreak structures for $> p_c$.

\section{\label{sec:methods}Model and Methods}

\subsection{\label{sec:percolation}The percolation model}

\begin{figure*}[tb]
    (a) 
    \raisebox{5ex}{
    \begin{minipage}[b]{0.3\textwidth}
    \includegraphics[width=0.45\columnwidth]{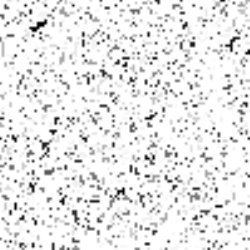}
    \includegraphics[width=0.45\columnwidth]{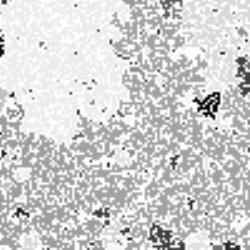}\\
    \includegraphics[width=0.45\columnwidth]{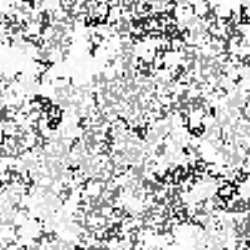}
    \includegraphics[width=0.45\columnwidth]{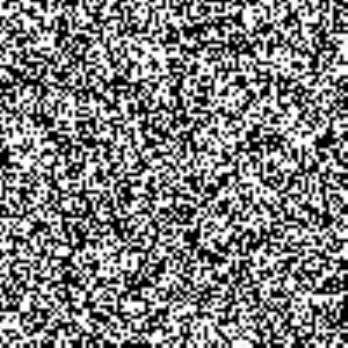}
    \end{minipage}}
    (b)\includegraphics[width=0.3\textwidth]{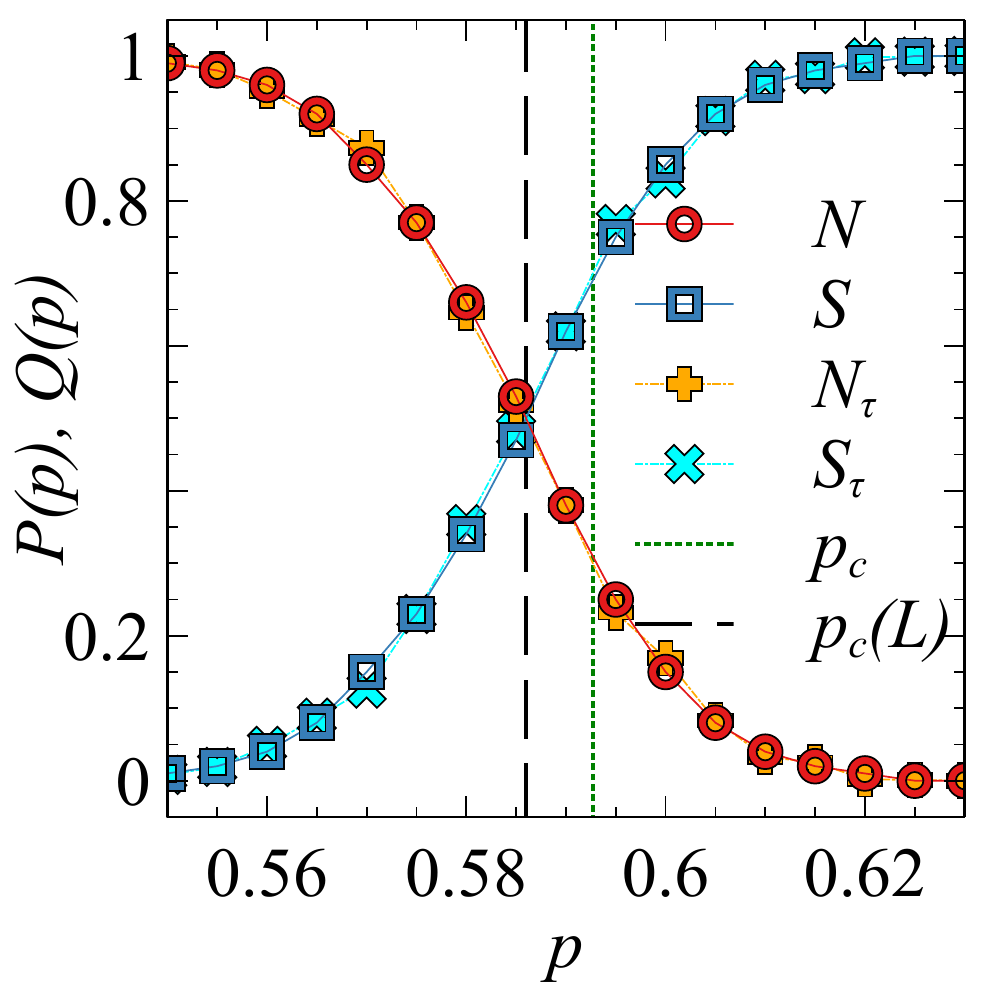} 
    (c)\includegraphics[width=0.3\textwidth]{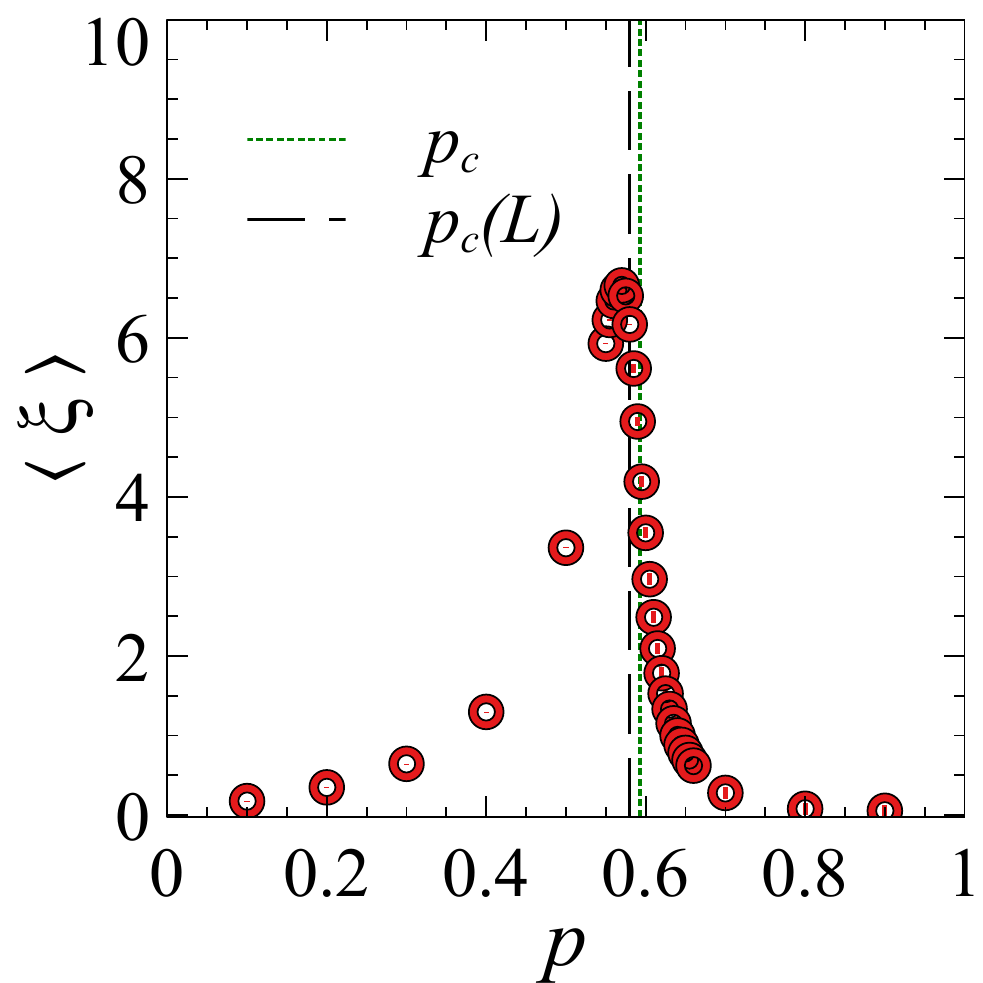} 
    \caption{
    (a) Examples of percolation clusters of size $L^2=100^2$, obtained for $p=0.2<p_c$, $0.6> p_c$ in the top row and $p=0.5$, i.e.\ close to $p_c$, in the bottom row. While individual clusters have been highlighted with different gray scales for the first three images, the bottom right image with $p=0.5$ shows all occupied sites in black only, irrespective of cluster identity. This latter representation is used below for the ML approach.
    (b) Percolation probabilities $P(p)$ and $Q(p)$ of having a spanning (blue open squares) / non-spanning (red open circles) cluster close to the percolation threshold for dataset $\mathcal{T} \cup \mathcal{V}$. The percolation probability of having a spanning (cyan crosses) / non-spanning (orange plus) cluster close to the percolation threshold for dataset $\tau$. 
    (c) Correlation length $\xi(p)$.
    In (b+c), the vertical lines indicate the estimates $p_c(100)\sim 0.585(5)$ (dotted) and $p_c$ (dashed).}
    \label{fig:percolation}
\end{figure*}

The percolation problem is well-known with a rich history across the natural sciences \cite{Broadbent1957a,Stauffer1991IntroductionTheory,Elliott1960EquivalenceFerromagnetism,Flory1953a,Derrida1985,Grimmett1989}. It  provides the usual statistical characteristics across a second-order transition such as, e.g., critical exponents, finite-size scaling, renormalization and universality \cite{Stauffer1991IntroductionTheory}. 
%
Briefly, on a percolation lattice of size $L \times L$, individual lattice sites $\vec{x}=(x,y)$, $x,y \in [1,L]$, are randomly occupied with \emph{occupation probability} $p$ such that the state $\psi$ of site $\vec{x}$ is $\psi(\vec{x})=1$ for occupied and $\psi(\vec{x})=0$ for unoccupied sites. 
We say that a connection between neighboring sites exists when these are side-to-side nearest-neighbors on the square lattice, while diagonal sites can never be connected. A group of these connected occupied sites is called a \emph{cluster}. Such a cluster then \emph{percolates} when it spans the whole lattice either vertically from the top of the square to the bottom or, equivalently, horizontally from the left to the right. Obviously, for $p=0$, all sites are unoccupied and no spanning cluster can exist while for $p=1$ the spanning cluster trivially extends throughout the lattice.
In Fig.\ \ref{fig:percolation}, we show examples of percolation clusters generated for various $p$ values.
The \emph{percolation threshold} is at $p=p_c(L)$, such that for $p< p_c(L)$ most clusters do not span while for $p > p_c(L)$ they do. This can be expressed as the \emph{percolation probability} $P(p)= \langle s_L(p)/L^2 \rangle$, where $s_L(p)$ gives the size of the (largest) percolating cluster for size $L$ and $\langle \cdot \rangle$ denotes an average over many randomly generated realizations. Similarly, we can define a probability of non-percolating, $Q(p)=\langle (L^2-s_L(p))/L^2 \rangle$ and $P(p) + Q(p) = 1$.
For an infinite system ($L\rightarrow\infty$), one finds the emergence of an infinite spanning cluster at $p_{c}=0.59274605079210(2)$. This estimate has been determined numerically evermore precisely over the preceding decades \cite{Jacobsen2014} while no analytical value is yet known \cite{Grimmett1989}.
Another quantity often used to characterize the percolation transition is the two-site correlation function $g(r)=\langle \psi(\vec{x}) \psi(\vec{x}+\vec{r})\rangle_{\vec{x},|\vec{r}|=r}$, where the $\langle \cdot \rangle_{\vec{x},|\vec{r}|=r}$ denotes the average over all $\vec{x}$ and directions $|\vec{r}|$. This $g(r)$ measures the probability to have two occupied sites separated by a distance $r$, in the same cluster \cite{Stauffer1991IntroductionTheory}. The associated correlation length $\xi$ is determined through 
$
 \xi = {\sqrt{{\sum_{r}r^2 g(r)}}}/{{\sum_{r}g(r)}}
$. 
In the infinite system $\xi$ diverges at $p_c$ as $|p-p_c|^{-\nu}$, where $\nu=4/3$ is the critical exponent, determining the universality class of the percolation problem \cite{Stauffer1991IntroductionTheory}. 

\subsection{\label{sec:data}Generating percolation states for training and validation}

In order to facilitate the recognition of percolation with image recognition tools of ML, we have generated finite-sized $L \times L$, with $L=100$, percolation states, denoted as $\psi_i(p)$, for the $31$ $p$-values $0.1, 0.2, \ldots$, $0.5, 0.55, 0.555, 0.556$, $\dots, 0.655, 0.66, 0.7$, $\ldots, 0.9$. For each such $p$, $N=10000$ different random $\psi_i(p)$ have been generated. 
Each state $\psi_i(p)$, $i=1, \ldots, N$, is of course just an array of numbers with $0$ denoting unoccupied and $1$ occupied sites. Nevertheless, we occasionally use for convenience the term ``image'' to denote $\psi_i(p)$.
The well-known Hoshen-Kopelman algorithm \cite{Hoshen1976a} is used to identify and label clusters from which we (i) compute $s(p)$, $g(r)$, and $\xi(p)$ as well as (ii) determine the presence or absence of a spanning cluster.
In Fig.\ \ref{fig:percolation} we show examples of percolation states generated for various $p$ values as well as the extracted $P_{100}$, $Q_{100}$, $p_{c}(100)$ and $\langle \xi(p)\rangle$ estimates. The different gray scales used in Fig.\ \ref{fig:percolation}(a) mark the different connected clusters. 
However, for the ML approach below, we shall only use the numerical values $0$ and $1$ corresponding to the state $\psi_i(p)$ \footnote{We have checked that our results do not change when using the gray scales. Furthermore, the full cluster information is already coded in the spatial distribution of these gray intensities, hence giving much information to the DNNs.}. This is visualized as the simple black and white version shown, e.g., for $p=0.5$ in Fig.\ \ref{fig:percolation} (a). 
From Figs.\ \ref{fig:percolation} (b+c), we note that $P(p)$ and $\xi(p)$ behave qualitatively as expected \cite{Stauffer1991IntroductionTheory}, with $P(p)\lesssim 1$ for $p < p_c$ and $P(p) \gtrsim 0$ for $p>p_c$ and $\xi(p)$ maximal near $p_c$. Clearly, $p_c(L=100)\sim 0.585(5) < p_c$. This latter behavior is as expected since $s_L(p) \leq s_{\infty}(p)$, i.e., a cluster that seemingly spans an $L \times L$ finite square might still not span on an infinite system.

We emphasize that in the construction, we took care to only construct states such that for each $p$, the number of occupied sites is exactly $N_\text{occ}= p \times L^2$ and hence $p$ can be used as exact label for the supervised learning approach. We note that $p= N_\text{occ} / L^2$ can therefore also be called the percolation \emph{density}. 
For the ML results discussed below, it will also be important to note that the spacing between $p$ values reduces when $p$ reaches $0.5$ with the next $p$ value given by $0.55$ and then $0.555$. Similarly, the $p$ spacing increases as $0.655$, $0.66$, $0.7$. We will later see that this results in some deviations from perfect classification/regression.
Last, we have also generated a similar test set with $L=200$. As the ML training cycles naturally take much longer, we have not averaged these over ten independent trainings. We find that our results do not change significantly when using this much larger data set and hence we will refrain from showing any result for these larger states in the following.

\subsection{\label{sec:supervised}Supervised ML strategy for phase prediction}

As discussed above, DL neural nets using the power of CNNs are among the preferred approaches when trying to identify phases in condensed matter systems \cite{Mehta2018, Dawid2022ModernWittek,Bedolla2020}. Here, we shall use a {\sc ResNet18} \cite{He2016DeepRecognition} network with $17$ convolutional and $1$ fully-connected layers, pretrained on the {\sc ImageNet} dataset \cite{Deng2009ImageNetAL}. As basis of  our ML implementation we use the {\sc PyTorch} suite of ML codes \cite{Paszke2019}.
We train the {\sc ResNet18} on the percolation of $310000$ states, using a $90\%$/$10\%$ split into training and validation data, $\mathcal{T}$ and $\mathcal{V}$, respectively; this corresponds to $N_\text{train}= 279000$ and $N_\text{val}= 31000$ samples, respectively.
%
We concentrate on two supervised ML tasks. First, we \emph{classify} percolation images according to (i) $p$, (ii) $\xi$ as well as (iii) spanning versus non-spanning. In the second task, we aim to predict $p$ and $\xi$ values via ML \emph{regression}. In both tasks, the overall network architecture remains identical, we just adapt the last layer as usual \cite{Neuralsupervised}. For the classification the output layers have a number of neurons corresponding to the number of classes trained, i.e., for the classification by density the ${\cal C}=31$ $p$-values given above, while for regression the output layer has only one neuron giving the numerical prediction.
However, the loss functions are different. 
Let $\mathbf{w}$ denote the set of parameters (weights) of the {\sc ResNet} and let $(\psi_i,\chi_i)$ represent a given image sample with $\chi_i$ its classification/regression target, i.e., classes $p$ or $\xi$, and also ${\chi}'_i$ the predicted values, $p'$ or $\xi'$. 
For classification of categorical data, the class names are denoted by a class index $c= 1, \ldots, {\cal C}$ and encoded as $\chi_{ck}=  1$  if $\chi_{c}=k$, $0$ otherwise.
Then, for the (multi-class) classification problem, we choose the usual cross-entropy loss function, $l_\mathrm{c}(\mathbf{w})=-\sum_{k=1}^{n}  \sum_{c=1}^{{\cal C}} \chi_{ck} \log{\chi'_{ck}}(\mathbf{w})+(1-\chi_{ck})\log[1-\chi'_{ck}(\mathbf{w})]$, where $n$ is the number of samples, i.e., either $N_\text{train}$ or $N_\text{val}$ \cite{Mehta2018}. We use the {\sc AdaDelta} optimizer \cite{AdaDelta} and find that a learning rate of $\ell_r=0.001$ produces good convergence \footnote{We also optimized other hyperparameters and tested other optimizers such as ADAM but found no significant performance improvement.}. Another good metric for the classification task is the accuracy $a$, which is the ratio of correct predictions over $n$. 
%
The loss function for the regression problem is given by the mean-squared error $l_\mathrm{r}(\mathbf{w})=\frac{1}{n} \sum_{k=1}^{n} [\chi_{k}- {\chi}'_k (\mathbf{w})]^2$ while $\ell_r$ and the optimizer remain the same.
When giving results for $l_\mathrm{c}$ and $l_\mathrm{r}$ below, we always present those after averaging over at least $10$ independent training and validation cycles, i.e., with a different initial split of the data into mini-batches. We use the notation $\langle l_\mathrm{c} \rangle$ and $\langle l_\mathrm{r}\rangle$ to indicate this averaging.
In the case of classification, we also represent the quality of a prediction by confusion matrices \cite{Mehta2018}. These graphically represent the predicted class labels as a function of the true ones in matrix form, with an error-free prediction corresponding to a purely diagonal matrix.
For comparison of the classification and regression calculations, we use in both cases a maximum number of $\epsilon_\mathrm{max}=20$ epochs. 

Our ML calculations train with a batch size of $256$ for classification and for regression. All runs are performed on NVIDIA Quadro RTX6000 cards.

\subsection{\label{sec:test-data}Generating test data sets}

We generate a test data set, $\tau$, of $1000$ states for each of the $31$ $p$-values, such that in total we have $N_\tau=31000$.
This test set is used to make all the confusion matrices given below. By doing this, we ensure that the performance of the trained DL networks is always measured on unseen data \cite{Dawid2022ModernWittek}.

In addition, we generate three special test data sets. These data sets have been constructed to allow testing for the existence of the spanning cluster.
The first special data set, $\tau_\text{sl}$, is made for the $27$ $p$-values $0.5, 0.55,0.555, \ldots, 0.66,0.7$ close to $p_c$ and again consists of $1000$ states $\psi_i(p)$ for each $p$. After generating each $\psi_i(p)$, we add a \emph{straight line} of occupied sites from top to bottom, while keeping $p$ constant by removing other sites at random positions. Obviously, every $\psi_i(p)$ in $\tau_\text{sl}$ therefore contains at least one spanning cluster by construction. 
As a consistency check to the performance of the ML networks, we also add two more $\psi_i$ without any connecting path for $p=0.1$ and $0.2$.

In the next set, $\tau_\text{rw}$, we start with the same $27$ $p$-values for a new set of $27000$ $\psi_i(p)$, but instead of the straight line, we add a directed \emph{random walk} from top to bottom. As before, we conserve the overall density $p$ of occupied sites. 
Hence, every samples in $\tau_\text{rw}$ is spanning. We again add two $\psi_i$ for $p=0.1$ and $0.2$ without the connected random path. 

\begin{figure}[tb]
    \centering%
    (a) \includegraphics[width=0.26\columnwidth]{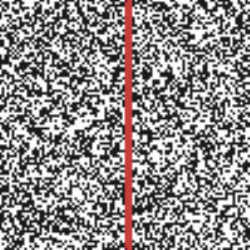}
    (b) 
    \includegraphics[width=0.26\columnwidth]{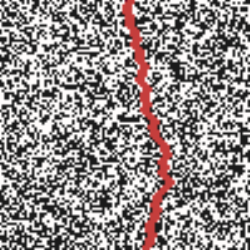}
    (c) \includegraphics[width=0.26\columnwidth]{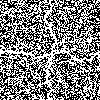}
    \caption{Examples of percolation images from the three special test sets $\tau_S$ with (a) a percolating straight line from top to bottom, (b) a percolating random path from top to bottom and (c) a "firebreak"-like cross of empty sites preventing percolation. For the sake of visibility, in (a+b) the connected path is highlighted in red. In all three cases, $p=0.5$.
    }
    \label{fig:special}
\end{figure}

Finally, the third special data set, $\tau_\text{fb}$, again contains $27 000$ lattices for the same previously mentioned $27$ $p$-values, but in each of the states we apply random \emph{firebreak} paths, horizontally and vertically, of unoccupied sites. This set is clearly non-spanning. Following the same logic as for $\tau_\text{sl}$ and $\tau_\text{rw}$, we add two spanning test samples above $p_c$ without the firebreak, namely, for $p=0.8$ and $0.9$. 
In all three cases, despite the modification in the lattices we ensure that $N_\text{occ}= p \times L^2$ and hence the occupation density is $p$. Examples of the three sets can be seen in Fig.\ \ref{fig:special}.

\section{\label{sec:results}Results}

\subsection{\label{sec:sup-density-classification}Classification of states labeled with density $p$}

\begin{figure}[tb]
    \centering%
    (a)\hspace*{-3ex}\lower0ex\hbox{\includegraphics[width=0.51\columnwidth]{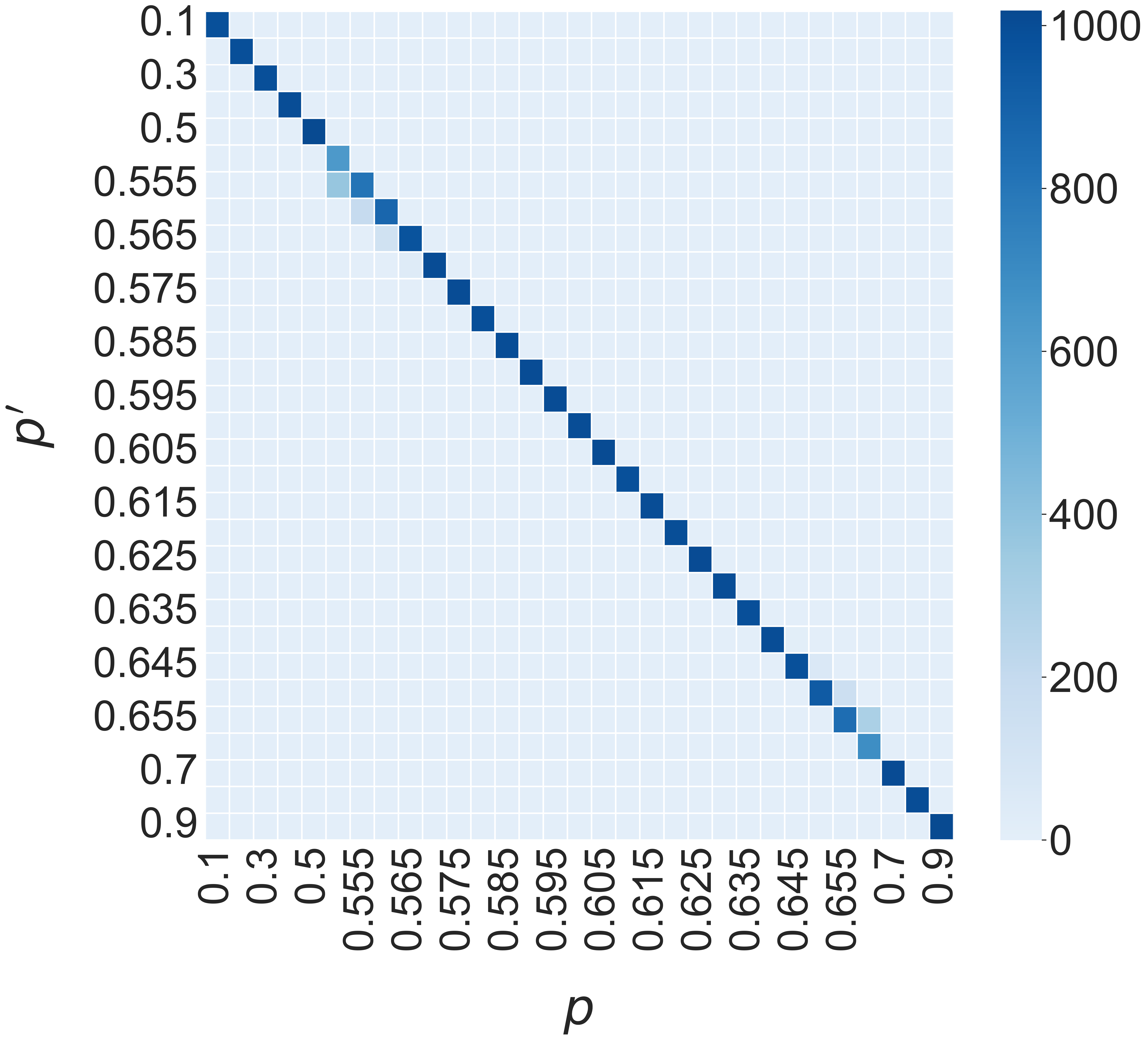}}
    (b)\hspace*{-3ex}\raise1ex\hbox{\includegraphics[width=0.44\columnwidth]{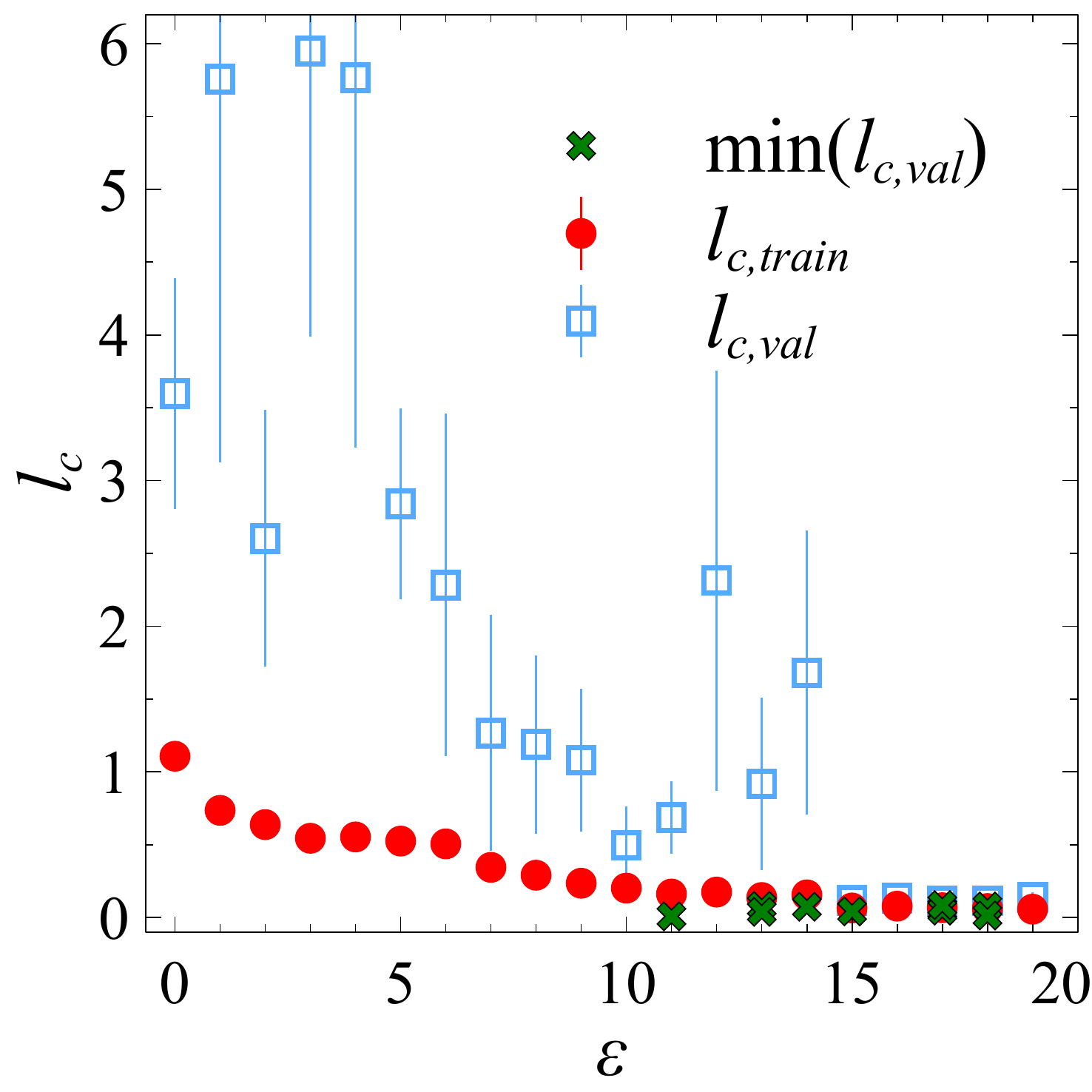}}
    
    \caption{ (a) Average confusion matrix for \textit{classification} according to $p$. The dataset used is the test data $\tau$ and the models used for predictions are those corresponding with a minimal $l_\text{c,val}$. True labels for $p$ are indicated on the horizontal axis while the predicted labels are given on the vertical axis. The color scale represents the number of samples in each matrix entry.
    (b) Dependence of losses $l_\mathrm{c,train}$ and $l_\mathrm{c,val}$ averaged over ten independent training seeds, on the number of epochs $\epsilon$ for \textit{classification} according to $p$. The squares (blue open) denote $l_\mathrm{c,train}$ while the circles (red solid) show $l_\mathrm{c,val}$. The green crosses indicate the minimal $l_\text{c,val}$ for each of the ten trainings.
    }
    \label{fig:ml-class-density-learningcurve}
\end{figure}

We use the density $p$ values as labels for the ML task of image recognition with the DL implementation outlined in section \ref{sec:supervised}. After ten trainings with all $310000$ images for $20$ epochs, we find on average a validation loss of $\langle l_\text{c,val} \rangle=0.052 \pm 0.009$ (corresponding to an accuracy of $\langle a_\text{c,val} \rangle =99.323\% \pm 0.003$). This is comparable to the very good image classification results shown on kaggle \cite{2012KaggleCats}. 
%
Fig.\ \ref{fig:ml-class-density-learningcurve}(a) gives the resulting averaged confusion matrix. The dependence of the training and validation losses, $\langle l_\mathrm{c,train} \rangle$ and $\langle l_\mathrm{c,val} \rangle$, respectively, on the number of epochs, $\epsilon$, is shown in Fig.\ \ref{fig:ml-class-density-learningcurve}(b). From the behavior of the loss functions, we can see that $\langle l_\mathrm{c,val} \rangle \geq \ \langle l_\mathrm{c,train} \rangle$ until $\epsilon=15$ after which both losses remain similar. This suggests that $\epsilon_\text{max}=20$ for our DL approach is indeed sufficient and avoids over-fitting.
%
Similarly, the confusion matrix is mostly diagonal with the exception of very few samples around the change of resolution in density, at $p \sim 0.555$ and $0.655$, as commented before in section \ref{sec:percolation}.

\subsection{\label{sec:sup-density-regression}Prediction of densities $p$ via regression}

\begin{figure}[tb]
    \centering%
    (a)\hspace*{-3ex}\raise1ex\hbox{\includegraphics[width=0.48\columnwidth]{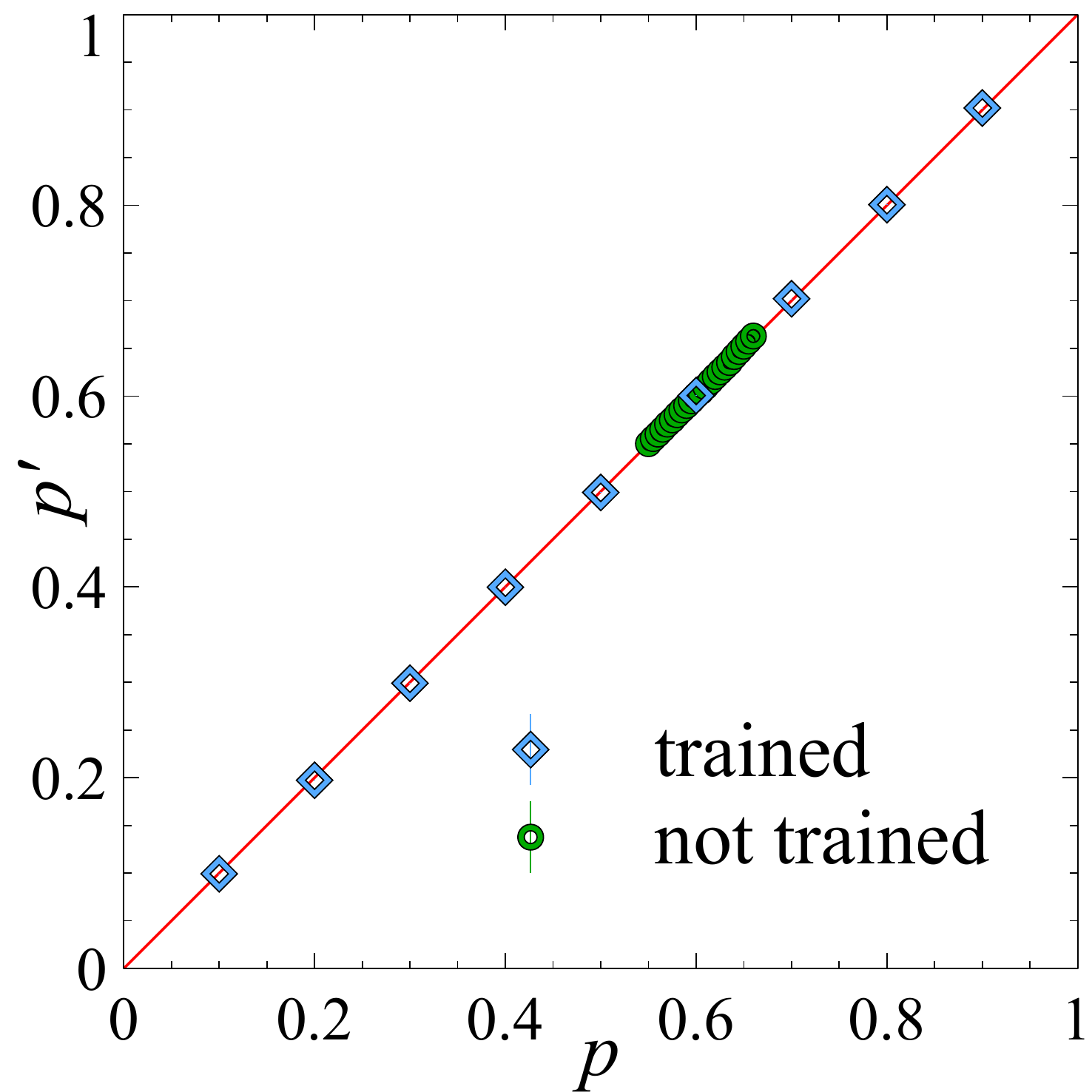}}\hfill%
    (b)\hspace*{-3ex}\raise0ex\hbox{\includegraphics[width=0.49\columnwidth]{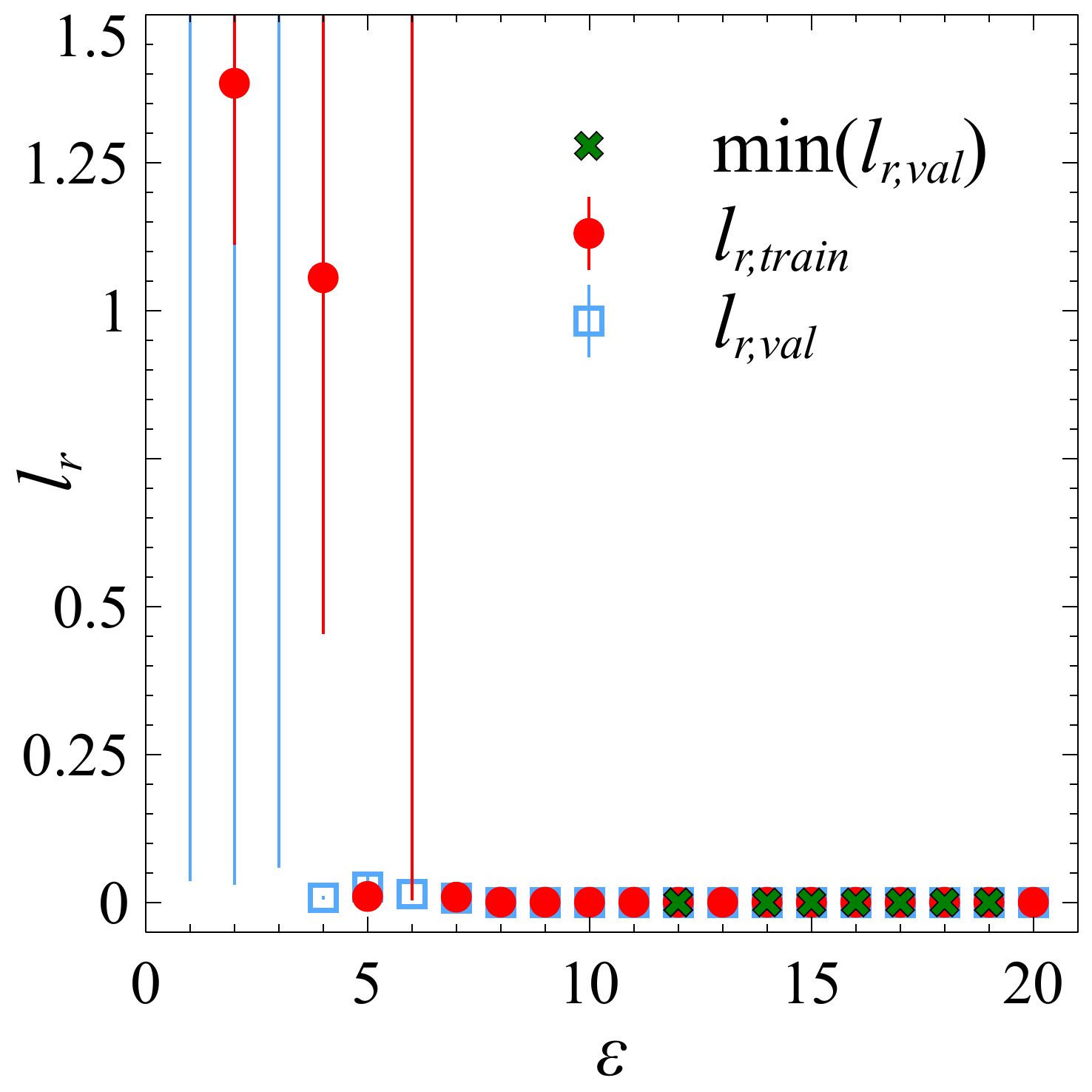}}    
    \caption{  (a) Average prediction curve obtained for \textit{regression} according to $p$ at the minimal $l_\text{r,val}$. The dataset used is the test data $\tau$ and the models used for predictions are those corresponding with a minimal $l_\text{c,val}$. The blue open squares denote $p$-values that have been used during the training and the green open circle shows $p$-values that were not trained.
    (b) Dependence of losses $l_\mathrm{r,train}$ and $l_\mathrm{r,val}$ averaged as in Fig.~\ref{fig:ml-class-density-learningcurve} on the number of epochs $\epsilon$ for \textit{regression} according to $p$. The squares (blue open) denote $l_\mathrm{r,train}$ while the circles (red solid) show $l_\mathrm{r,val}$. The green crosses show the minimal $l_\text{r,val}$ for each of the ten trainings.
    }
    \label{fig:ml-density-regression}
\end{figure}

For the regression problem, we train the {\sc ResNet18} only for the nine evenly spaced densities $p= 0.1, 0.2, \ldots, 0.9$. After training and validation with $\mathcal{T}$ and $\mathcal{V}$, respectively, we examine the states in $\tau$ and predict their $p$ values. 
In Fig.\ \ref{fig:ml-density-regression}, we present the results with (a) indicating the fidelity of the predictions for each $p$-value and (b) showing good convergence of the losses $l_\mathrm{r,train}$ and $l_\mathrm{r,val}$. 
Clearly, the regression works very well for the nine trained $p$-values $p=0.1, \ldots, 0.9$  as well as the untrained values $0.55, 0.555, \ldots, 0.0.655, 0.66$ close to $p_c(100)$.
After reaching $\epsilon=20$, we find that $\text{min}_{\epsilon} [ \langle l_\mathrm{r,train} \rangle] = 0.0003 \pm 0.0002$ and $\text{min}_{\epsilon} [\langle l_\mathrm{r,val} \rangle] = (6.2 \pm 1.2) \times 10^{-5}$.

Therefore we conclude that our CNN performs well for classification and regression tasks while $\mathcal{T}$, $\mathcal{V}$, and $\tau$ present appropriately structured data sets for these ML tasks in terms of data size.

\subsection{\label{sec:sup-correlation-classification}Classification with correlation length $\xi$}

We now turn our attention to studying image recognition when using the correlation lengths $\xi$, instead of $p$, as labels for the $\psi_i(p)$ states. One way to do this is to use $\langle \xi(p) \rangle$ as label. While for the classification by $p$  the label value was identical to the actual density $p$ of a given state, now each state is labeled by $\langle \xi (p)\rangle$. This means that the actual $\xi$ of the state might be different from the label assigned. Since $\langle \xi(p)\rangle$ can be uniquely identified by $p$, this strategy should be in fact equivalent to the previous situation and the CNN should give us similar classification results.
\begin{figure}[tb]
    \centering%
    (a)\hspace*{-3ex}\raise0ex\hbox{\includegraphics[width=0.51\columnwidth]{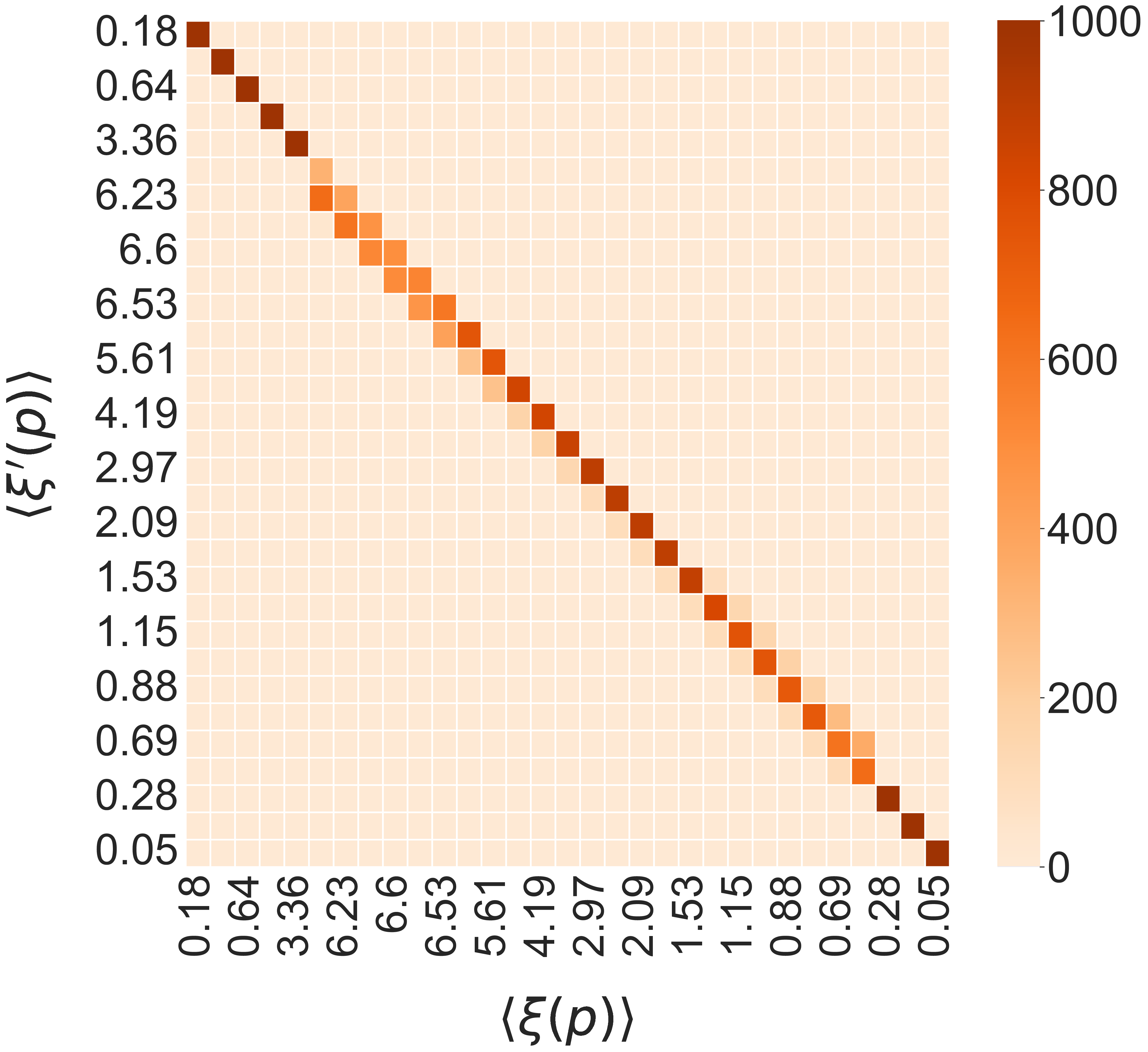}}
    (b)\hspace*{-3ex}\raise1ex\hbox{\includegraphics[width=0.44\columnwidth]{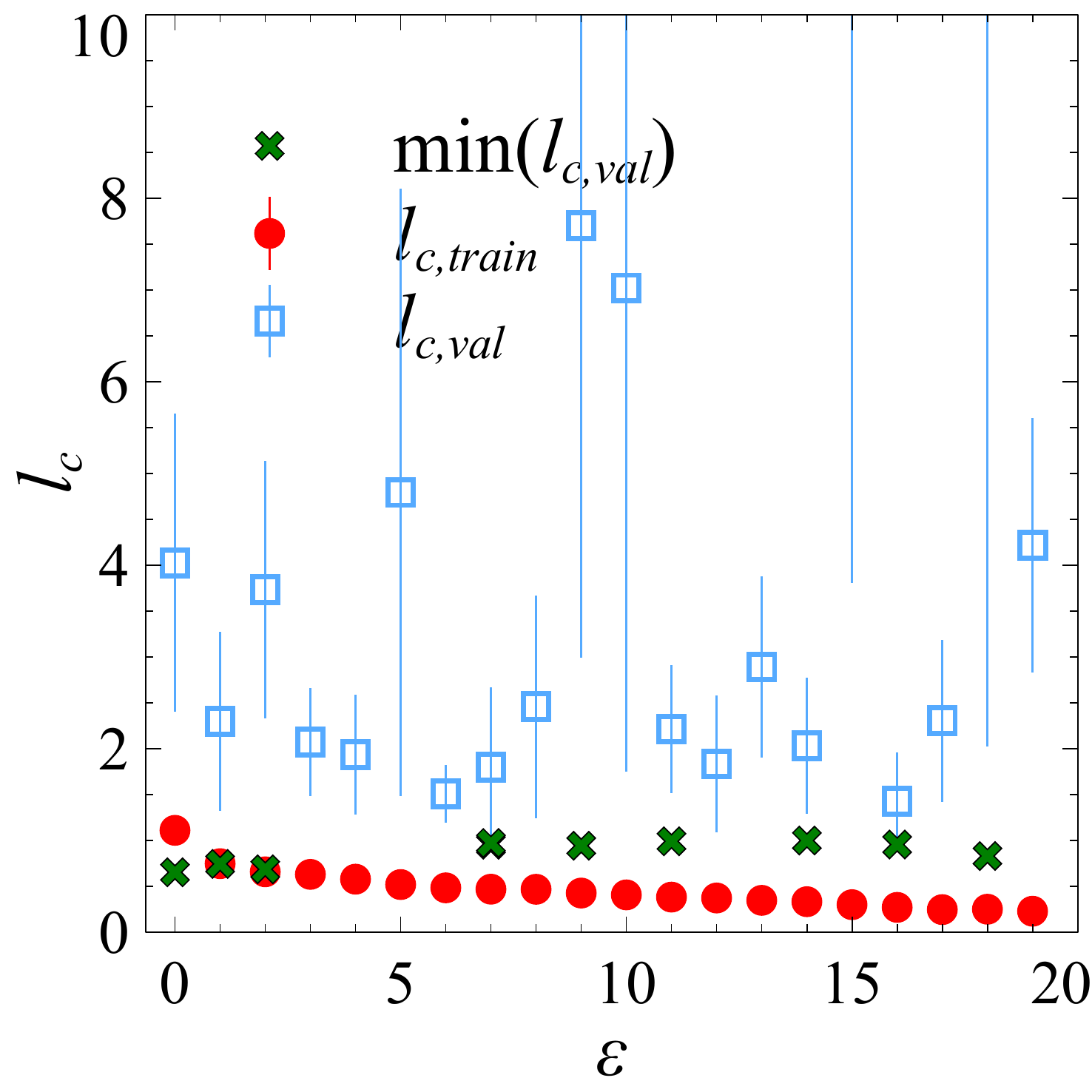}}
    \caption{(a) Average confusion matrix for \textit{classification} according to  $\langle \xi \rangle$. The dataset used is the test data $\tau$ and the models used for predictions are those corresponding with a minimal $l_\text{c,val}$.
    (b) Dependence of losses $l_\mathrm{c,train}$ and $l_\mathrm{c,val}$ on the number of epochs $\epsilon$ for classification according to $\langle \xi \rangle$. We follow the same convention as in Fig.\ \ref{fig:ml-class-density-learningcurve} and Fig.\ \ref{fig:ml-correlation-classification-pdf}.}
    \label{fig:ml-correlation-classification}
\end{figure}
The results of such a classification are shown in Fig.\ \ref{fig:ml-correlation-classification} where similarly to Fig.\ \ref{fig:ml-class-density-learningcurve} we present in (a) the average confusion matrix for the $31$ $\langle\xi(p)\rangle$ values (cf.\ also Fig.\ \ref{fig:percolation}) and in (b) the evolution of losses during the training. We find a validation loss of $\min_{\epsilon}[\langle l_\text{c,val} \rangle]=0.38 \pm 0.07$ (corresponding to a maximal accuracy of $\text{max}_{\epsilon}[\langle a_\text{c,val} \rangle]=87.12\% \pm 0.05$) and a highly diagonal confusion matrix, with only a small deviation that can be linked to the change in resolution in our data set above $p=0.5$. 

\begin{figure}[tb]
    (a)\includegraphics[width=0.44\columnwidth]{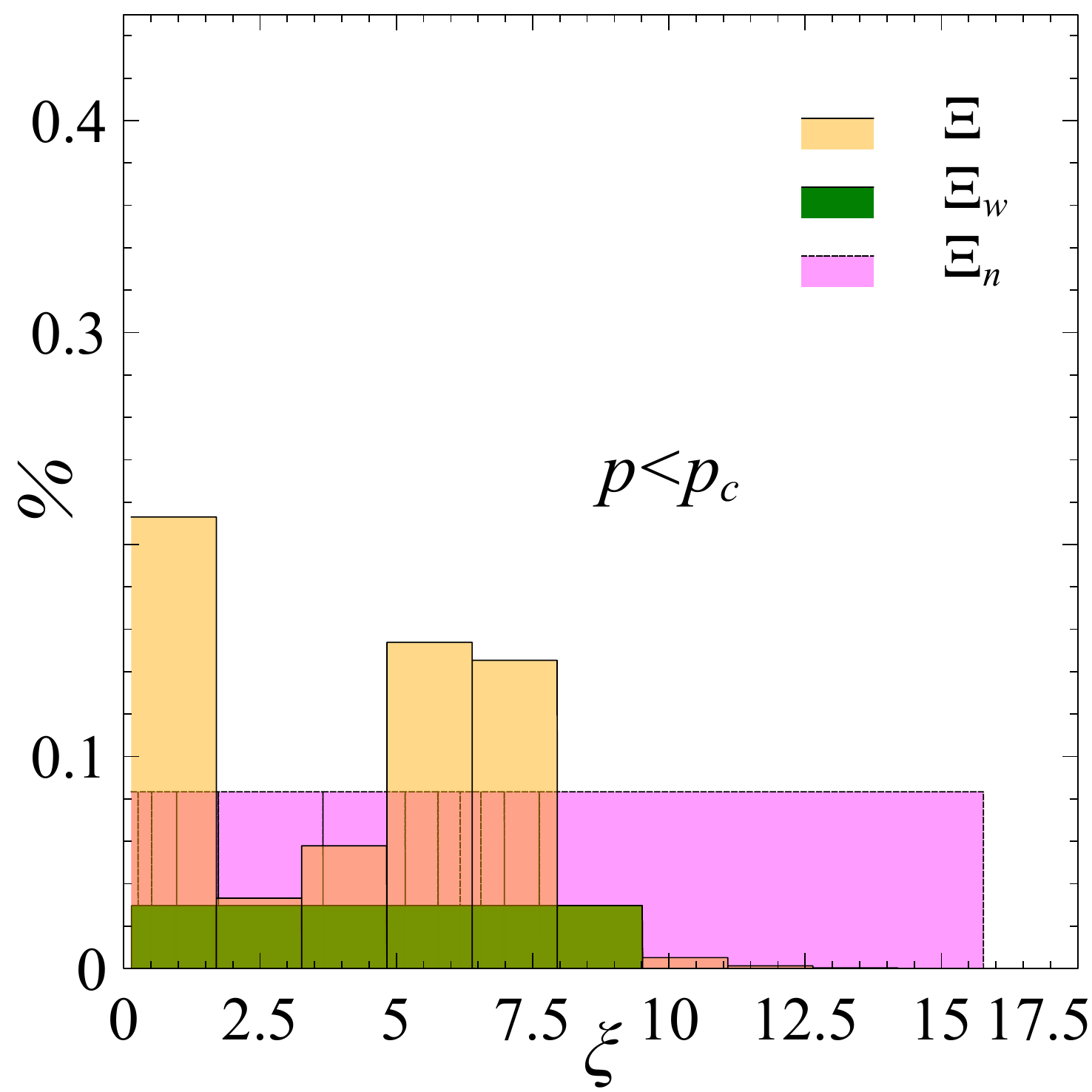}
    (b)\includegraphics[width=0.44\columnwidth]{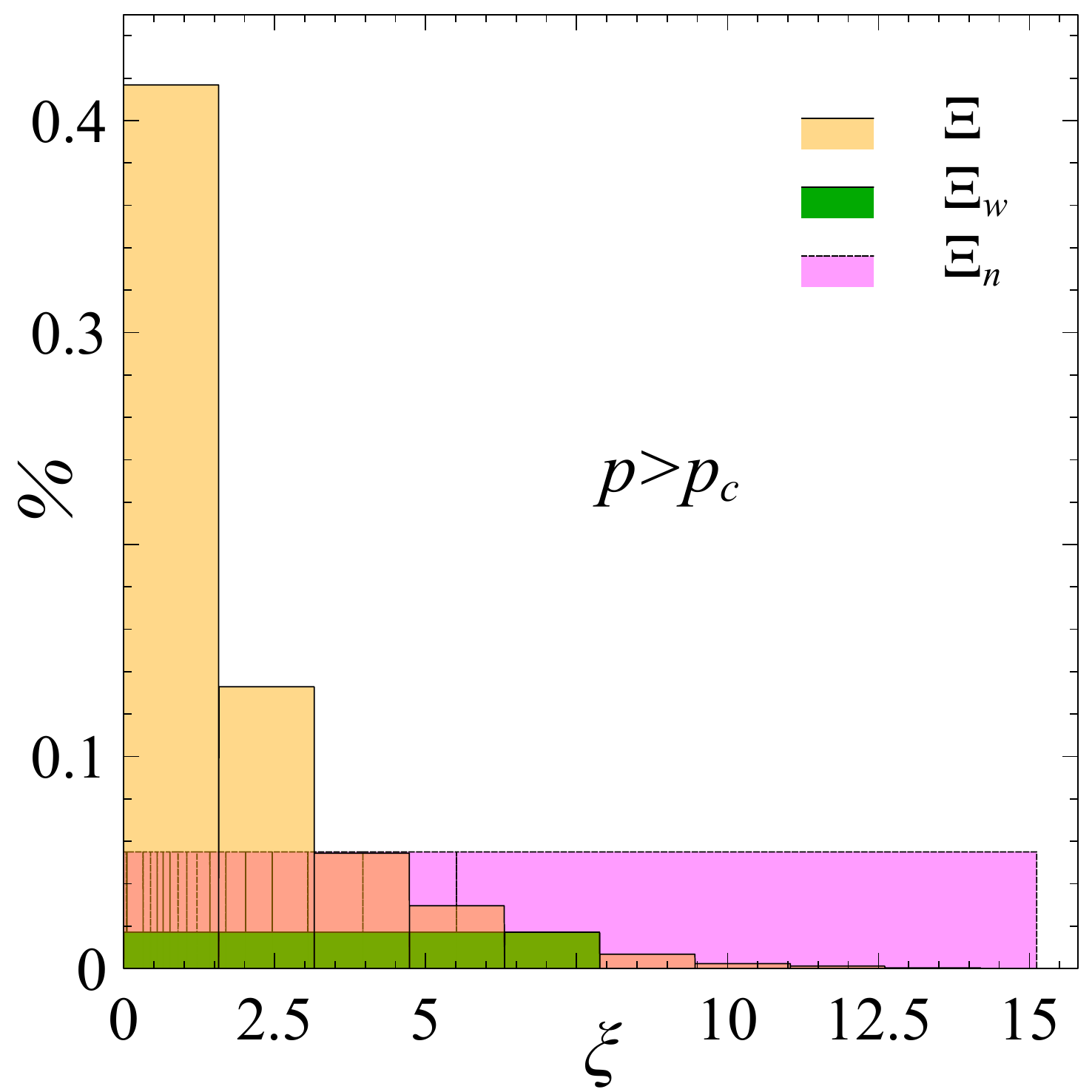}
    \caption{Probability distributions for correlation lengths $\xi$ when (a) $p<p_c$ (with $12$ $p$-values) and $p>p_c$ ($18$ $p$-values) with unbalanced $\Xi$ and the balanced counterparts $\Xi_n$ and $\Xi_w$ denoted by yellow, magenta and green, respectively. In each case, the distributions are normalized relative to the total number of $\xi$'s in each set, i.e.\ for (a) $120000$ in $\Xi$ and $\Xi_n$ and $6 \time 3560= 21360$ in $\Xi_w$ while for (b) there are $180000$ in $\Xi$ and $\Xi_n$ and $5\times 3077=15385$ in $\Xi_w$.
    }
    \label{fig:correlation-classification-distribution}
\end{figure}

One might wish to interpret the above classification with $\langle \xi(p) \rangle$ as a success of the ML approach. However, let us reemphasize that it is fundamentally equivalent to simply changing labels while keeping the direct connection of the labels with $p$ unaltered. 
We now wish to obtain a classification of states via their $\xi$'s which is independent of the $p$'s. In Fig.\ \ref{fig:correlation-classification-distribution} we show the distribution $\Xi$ of the $\xi$'s in $\mathcal{T} \cup \mathcal{V}$.
Clearly, the number of small $\xi$ values is larger than the number of $\xi$ values close to the maximal value of $\text{max}[\xi]=15.771$ (cp.\ Fig.\ \ref{fig:percolation}(c)). Hence simply using each $\xi$ as label for the corresponding $\psi_i$ would result in a biased dataset.
We therefore reorganize the $\mathcal{T} \cup \mathcal{V}$ data set. This can be done in two ways. 
For the first reorganization we create bins a constant \emph{number} of $10000$ samples in each bin. We call this dataset $\Xi_n$. This results in a varying bin width. 
The second way to reorganize the data set is to keep the bin width constant while restricting the \emph{number} of samples in each bin.  We shall denote this reorganization as $\Xi_w$.
Since $\xi(p)$ is non-monotonic in $p$, we split the reorganization into the case (i) $p < p_c$ with and (ii) $p>p_c$.
We emphasize that the reorganized data sets consist of the same states as in $\mathcal{T} \cup \mathcal{V}$ but now have different labels according to the bin labels for $\Xi_w$ and $\Xi_n$.
Furthermore, there is now no longer any direct connection of the new labels to the original $p$ densities.

\begin{figure*}[tb]
    \centering%
    \begin{minipage}{0.49\textwidth}
        \hspace*{5ex} $\Xi_w$, $p<p_{c}(L)$ \\
        (a.1)\hspace*{-1ex} \includegraphics[width=0.47\columnwidth]{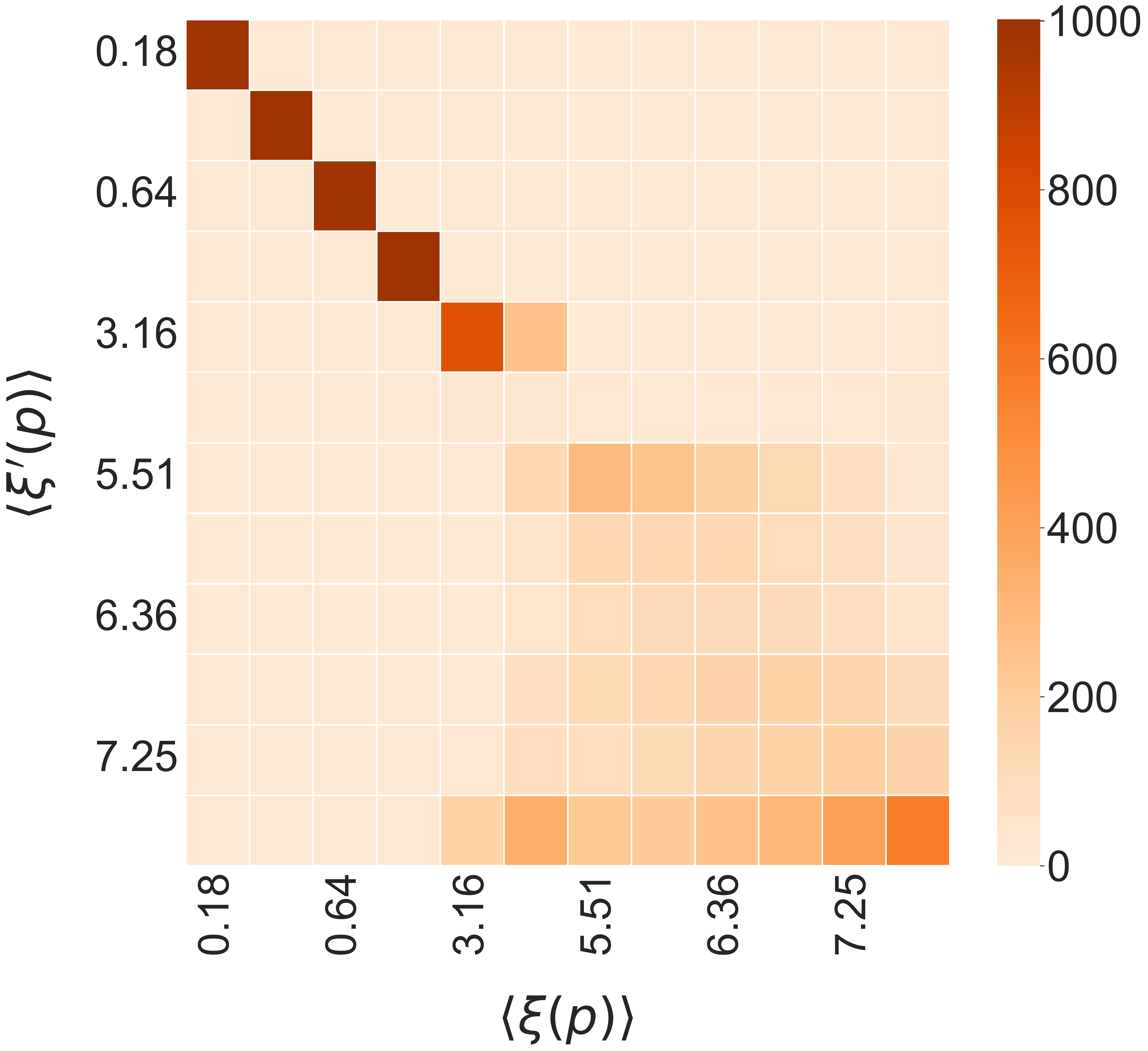} 
        \includegraphics[width=0.43\columnwidth]{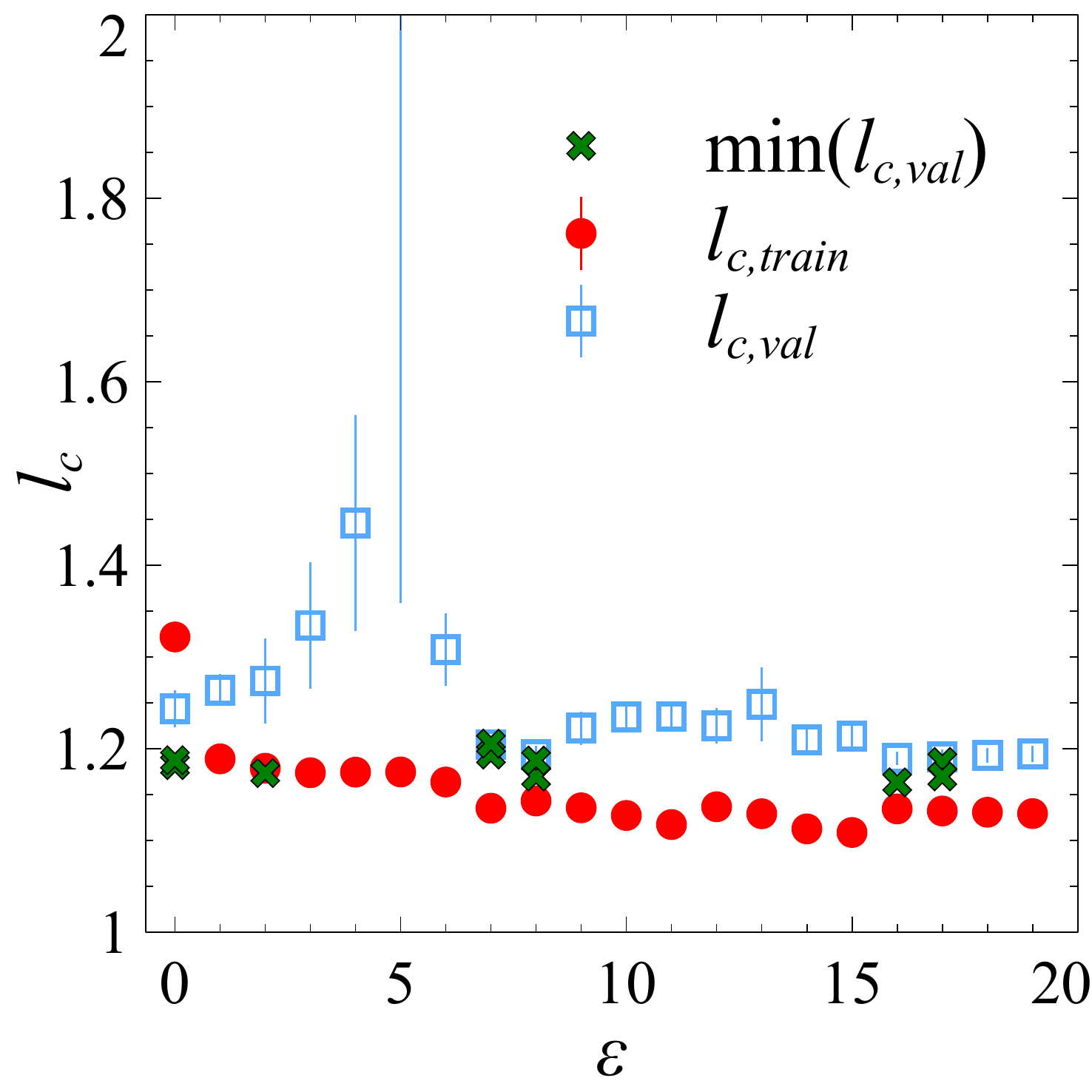}\\
        \hspace*{5ex} $\Xi_n$, $p<p_{c}(L)$\\
        (a.2)\hspace*{-1ex} \includegraphics[width=0.47\columnwidth]{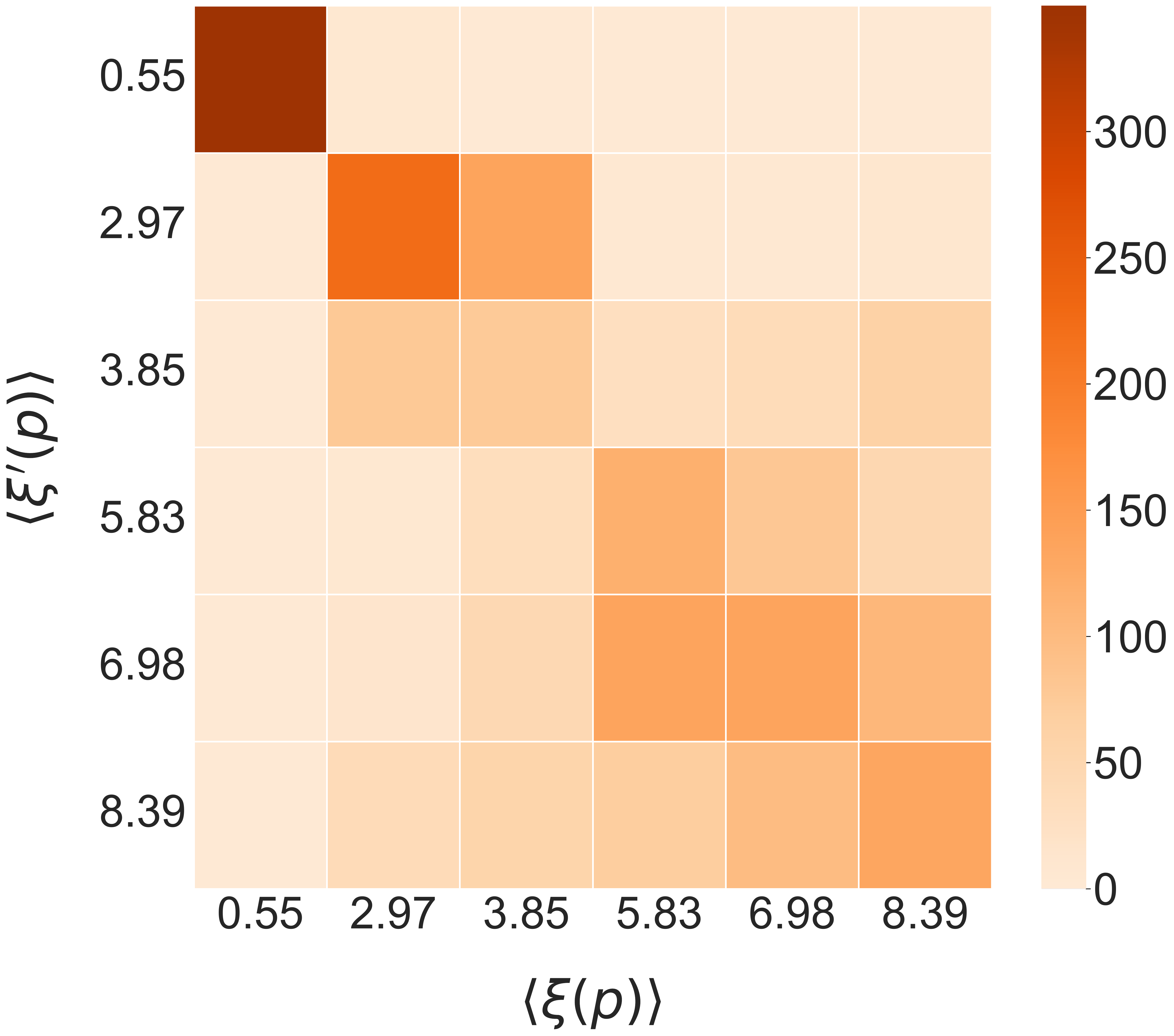} 
        \includegraphics[width=0.43\columnwidth]{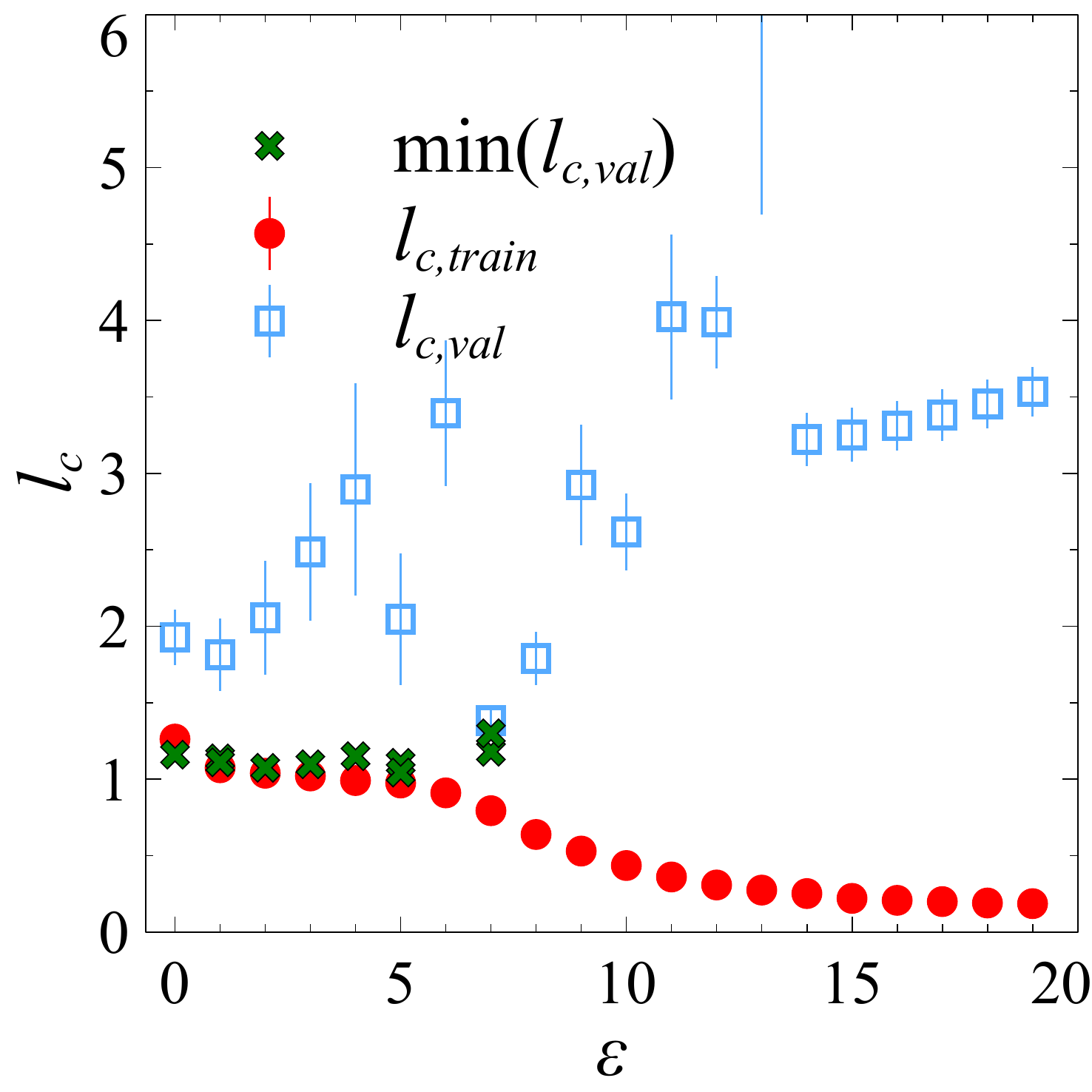}
    \end{minipage}
    %
    \begin{minipage}{0.49\textwidth}
        \hspace*{5ex} $\Xi_w$, $p>p_{c}(L))$ \\
        (b.1)\hspace*{-1ex} \includegraphics[width=0.47\columnwidth]{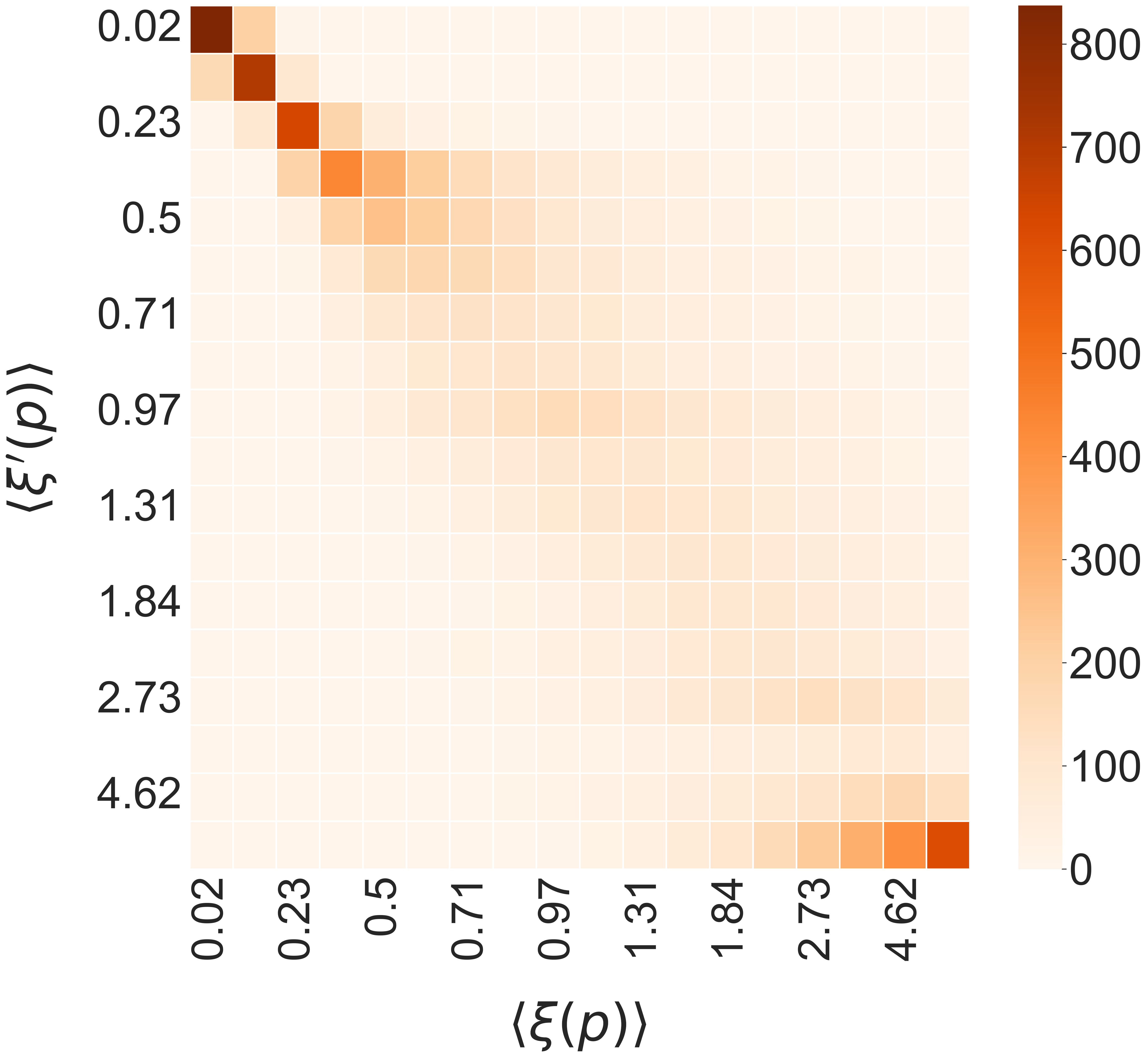}
        \includegraphics[width=0.43\columnwidth]{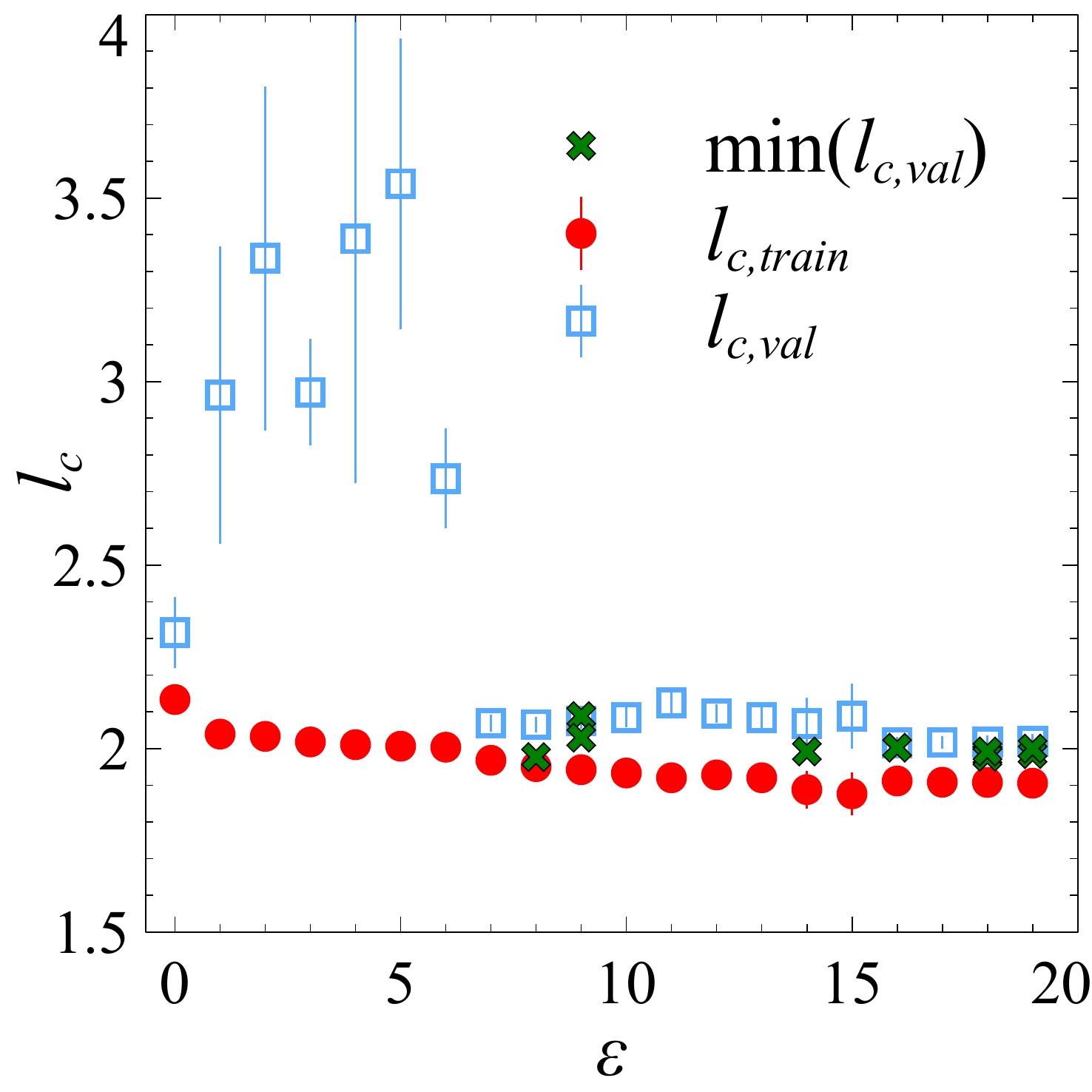}\\
        
        \hspace*{5ex} $\Xi_n$, $p>p_{c}(L)$ \\
        (b.2)\hspace*{-1ex} \includegraphics[width=0.47\columnwidth]{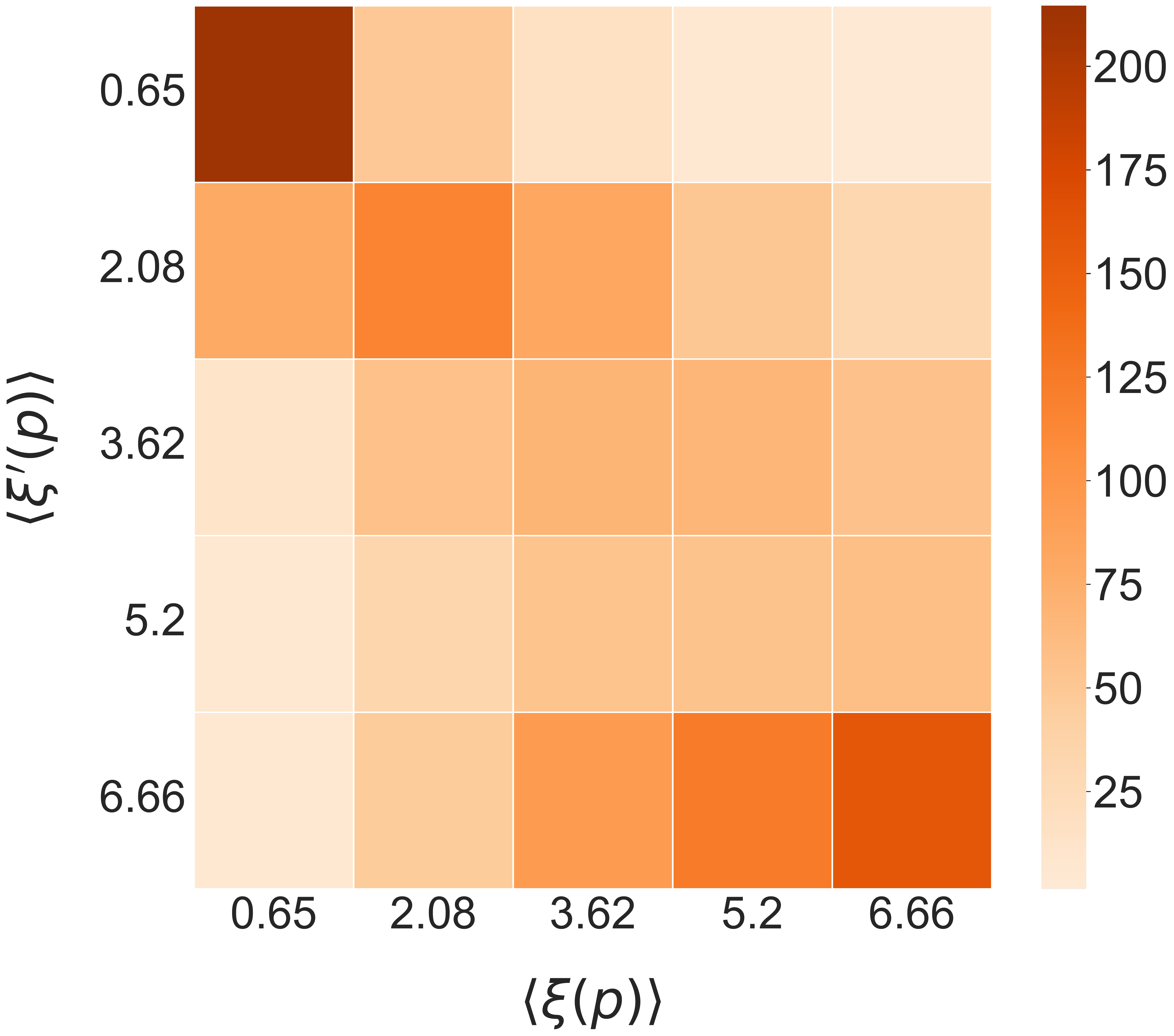}
        \includegraphics[width=0.43\columnwidth]{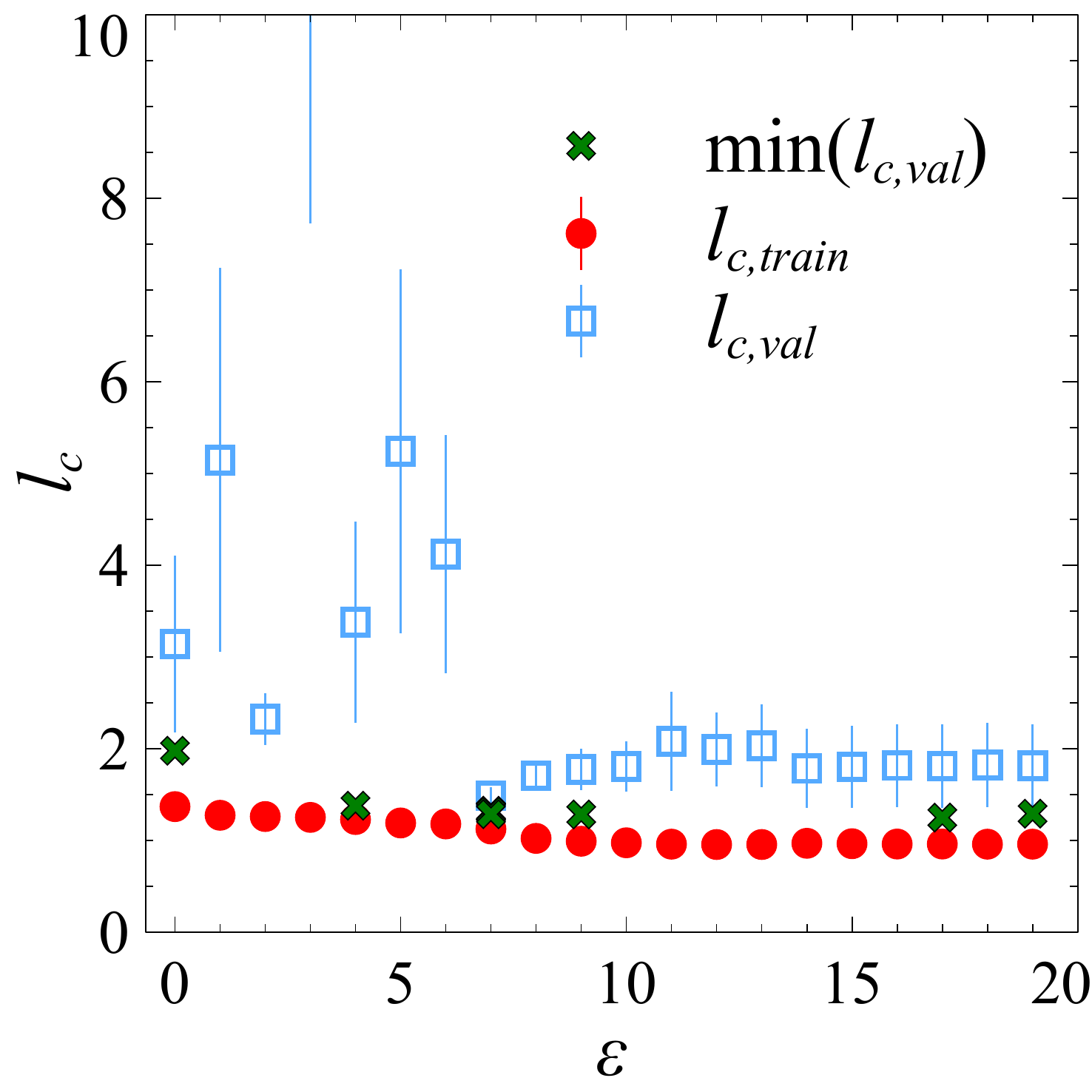}\\
       
    \end{minipage}
    \caption{\label{fig:ml-correlation-classification-pdf}
    Confusion matrices and losses $l_\mathrm{r,train}$ and $l_\mathrm{r,val}$ for the classification results when using the correlation-function-relabeled $\Xi_w$ and $\Xi_n$ data sets. The left column (a) shows the case $p < p_c$ while the right column (b) gives the outcome for $p>p_c$. The upper row corresponds to $10000$ states for each class with $12$ classes for $p< p_c$ and $18$ at $p > p_c$ for $\Xi_w$. In the lower row, there are $3560$ states for the $6$ classes when $p< p_c$ and $3077$ states for $5$ classes when $p > p_c$.}
\end{figure*}

In Fig.\ \ref{fig:ml-correlation-classification-pdf}, we plot the resulting confusion matrices and losses.
We see that the classification for $\Xi_w$ and $\Xi_n$ only results in large diagonal entries in the confusion matrices for small correlation lengths labels $\xi$. Overall, the classification for $\Xi_w$ is somewhat better than for $\Xi_n$ when away from $p_c(L)$. We attribute this to the larger number of states for the $\Xi_w$ for $p < p_c(L)$. 
Still, with overall $62.6\%$ and $55.1\%$ of states misclassified for $\Xi_w$ and $\Xi_n$, respectively, it seems clear that classification for correlation lengths must be considered unsatisfactory.

\subsection{\label{sec:sup-correlation-regression}Regression with correlation length $\xi$}

\begin{figure}[tb]
    \centering%
    (a)\hspace*{-3ex}\raise1ex\hbox{\includegraphics[width=0.48\columnwidth]{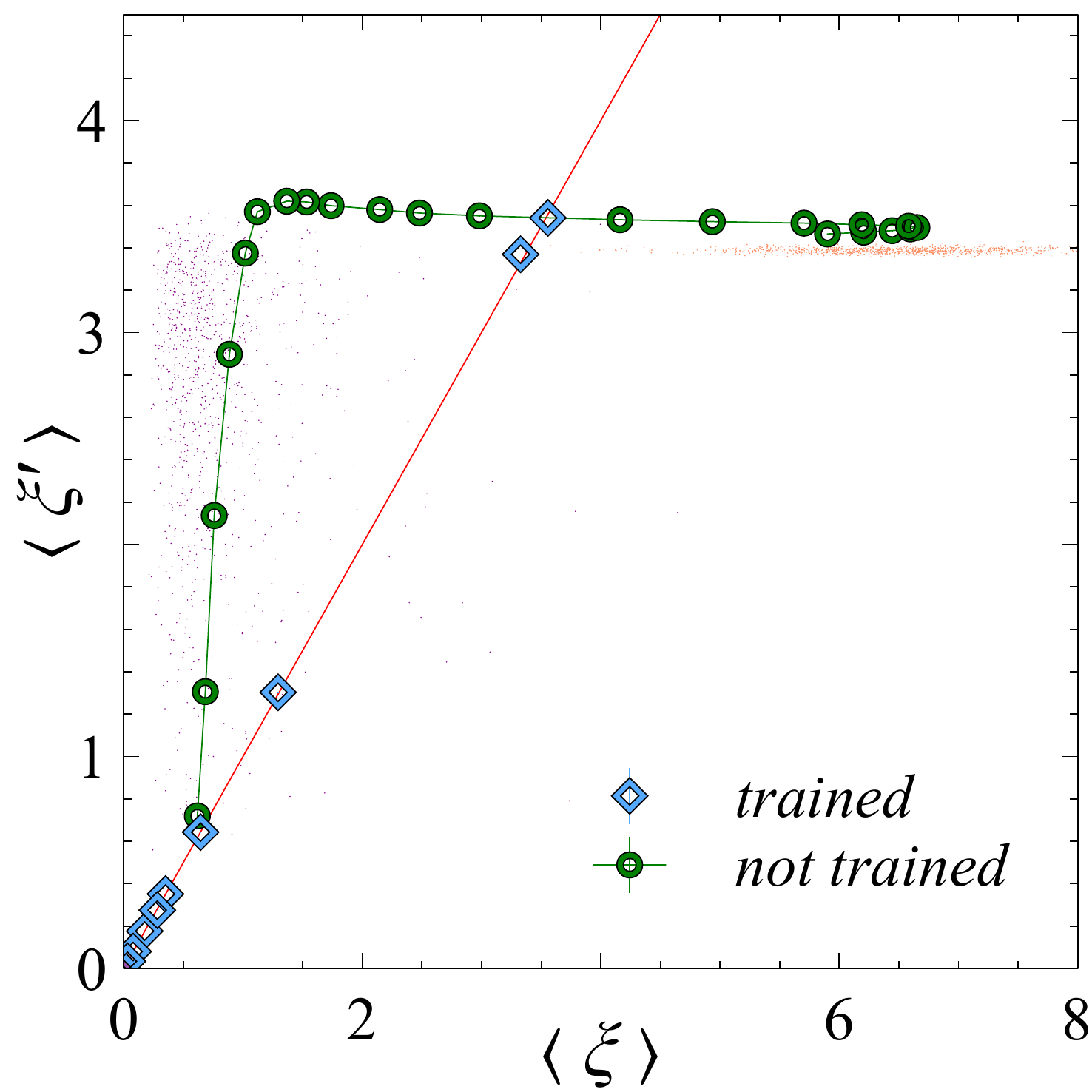}}
    (b)\hspace*{-3ex}\raise0ex\hbox{\includegraphics[width=0.49\columnwidth]{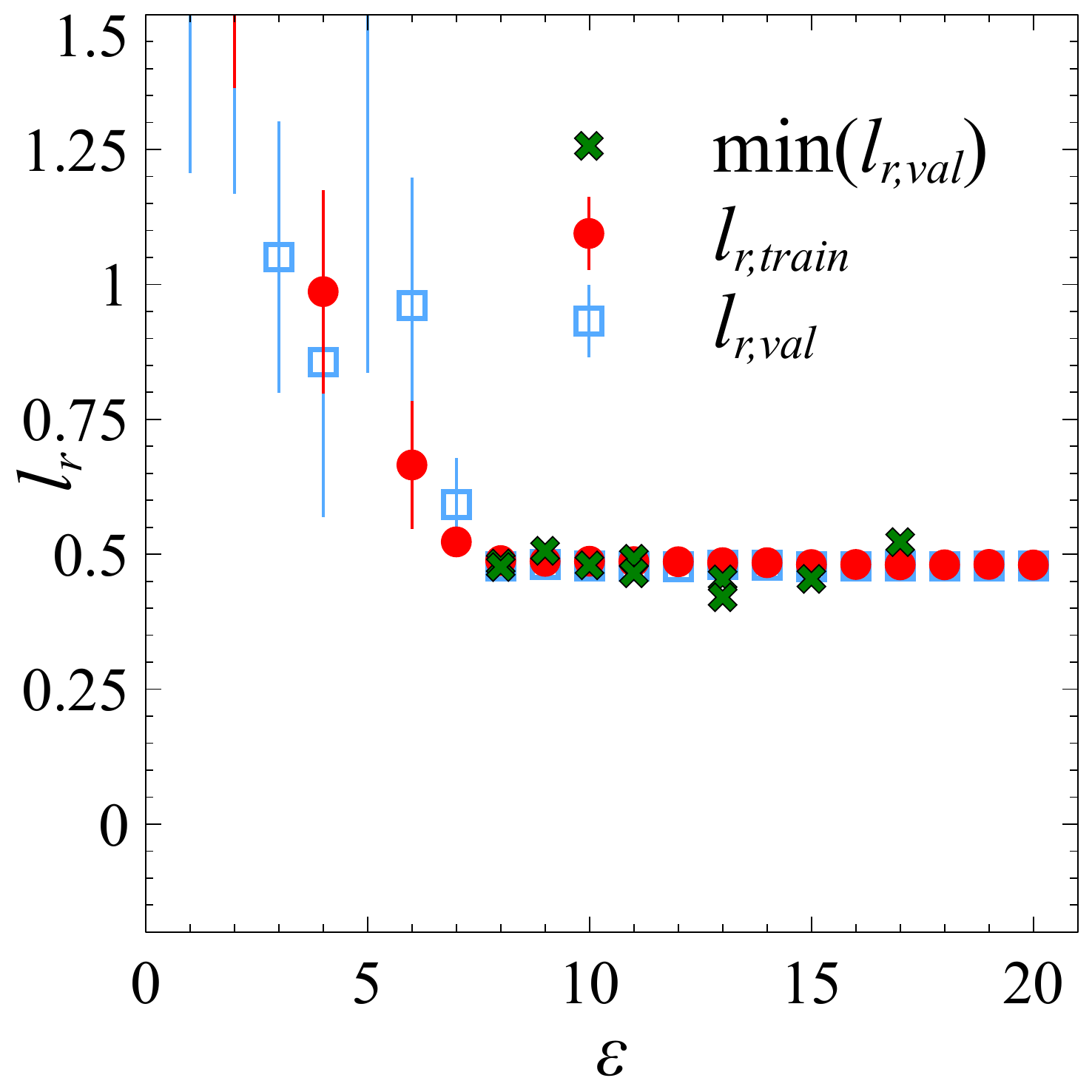}}
    \caption{ (a)  Average predictions for \textit{regression} according to  $\xi$. The dataset used is the test data $\tau$ and the models used for predictions are those corresponding with a minimal $l_\text{c,val}$. 
    (b) Dependence of losses $l_\mathrm{r,train}$ and $l_\mathrm{r,val}$ on the number of epochs $\epsilon$ for regression according to $\xi$. We follow the same convention as in Fig.\ \ref{fig:ml-density-regression}. }
    \label{fig:ml-correlation-regression}
\end{figure}

For the regression task with $\xi$, we proceed analogously to section \ref{sec:sup-density-regression}. Again, we train the CNN for the individual correlation length $\xi_i$ corresponding to each $\psi_i \in \mathcal{T}$ for the nine densities $p=0.1, \ldots, 0.9$. We then compute the predictions of $\xi_i$ for all $31$ densities in $\tau$. 
The results are shown in Fig.\ \ref{fig:ml-correlation-regression}.
We find that the network architecture which previously predicted the density quite accurately is now struggling to correctly predict $\xi$. A structure seems to exist in the predictions. By looking closely we notice that the network make use of the density for its predictions. Furthermore, by plotting the correlation length according to the density we retrieve the plot of $\xi$ Fig.\ \ref{fig:percolation}.

\subsection{\label{sec:sup-connectivity-total}Classification with the spanning or non-spanning properties}
 
As discussed earlier, the hallmark of the percolation transition is the existence of a spanning cluster which determines whether the system is percolating or not \cite{Stauffer1991IntroductionTheory}. In the previous section, our DL approach has classified according to $p$ or $\xi$ values without testing whether spanning clusters actually exist. 
We now want to check this and label all states according to whether they are spanning or non-spanning. From Fig.\ \ref{fig:percolation}, it is immediately clear that for finite-sized systems considered here, there are a non-negligible number of states with appear already spanning even when $p < p_c$ and, vice versa, are still non-spanning when $p > p_c$. Furthermore, we note that for such $L$, the difference between $p_c$ and $p_c(L)$ is large enough to be important and we hence use $p_c(L)$ as the appropriate value to distinguish the two phases.

\begin{figure}[tb]
    \centering%
    (a)\hspace*{-2ex}\raise3ex\hbox{\includegraphics[width=0.46\columnwidth]{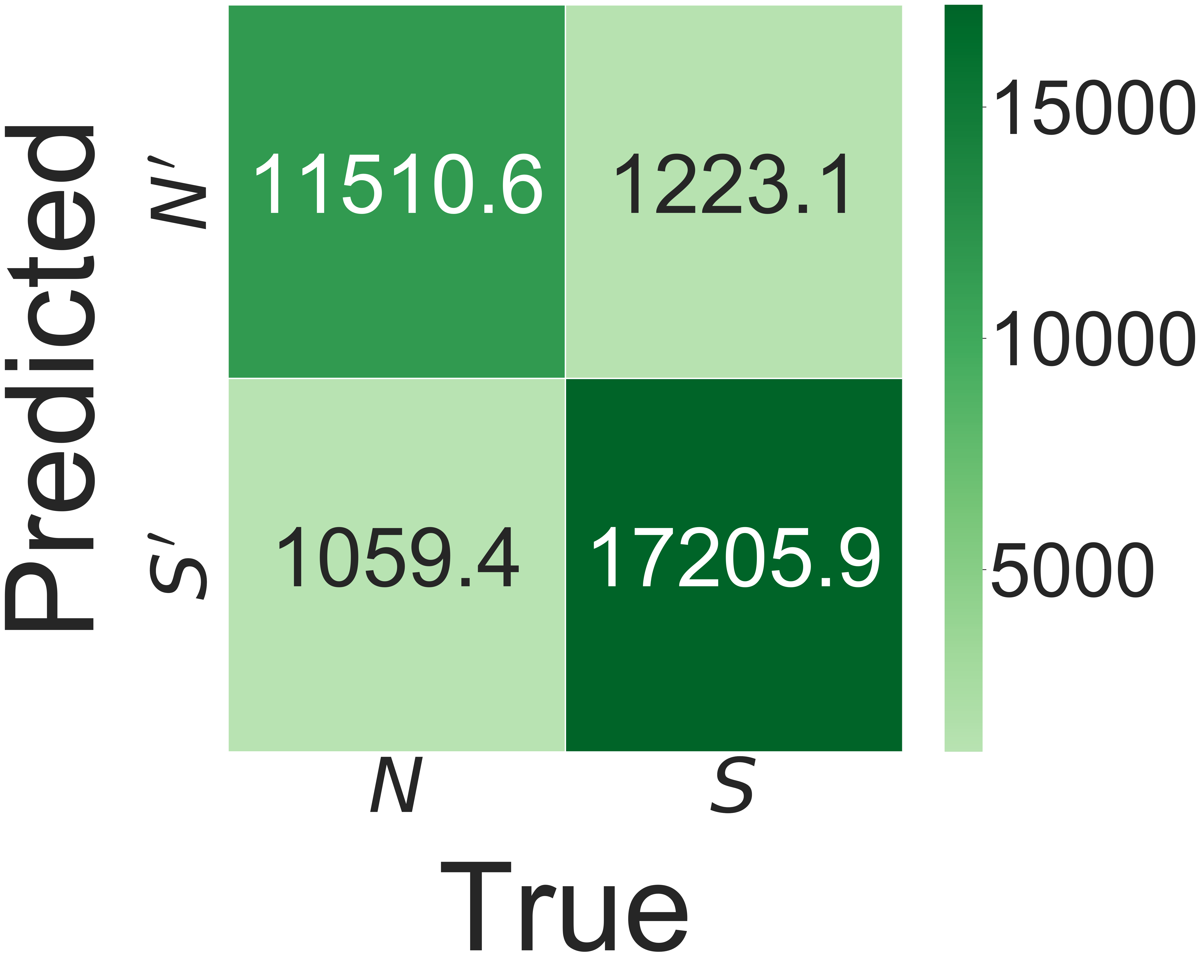}}%
    \hfill%
    (b)\hspace*{-3ex}\raise0ex\hbox{\includegraphics[width=0.48\columnwidth]{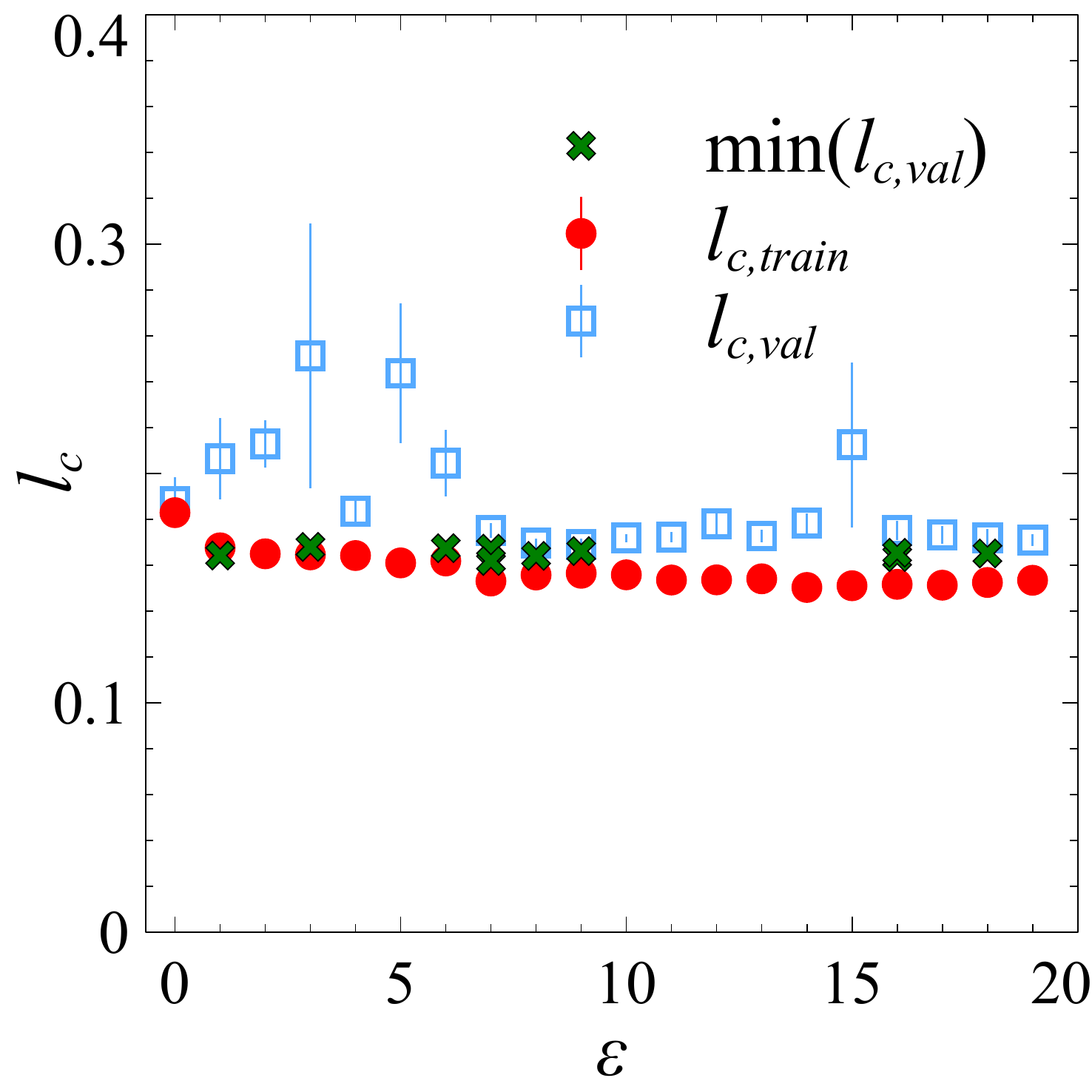}}
    \caption{(a) Average confusion matrix for \textit{classification} according to  spanning/non-spanning. The dataset used is the test data $\tau$ and the models used for predictions are those corresponding with a minimal $l_\text{c,val}$. The true labels for $N$ and $S$, are indicated on the horizontal axis while the predicted labels are given on the vertical axis. (b) Dependence of losses $l_\mathrm{c,train}$ and $l_\mathrm{c,val}$ on the number of epochs $\epsilon$ for classification according to spanning/non-spanning. Again, we follow the same convention as for Figs.\ \ref{fig:ml-class-density-learningcurve}, \ref{fig:ml-correlation-classification-pdf}, and \ref{fig:ml-correlation-classification}
    }
    \label{fig:ml-connectivity-total}
\end{figure}

Figure \ref{fig:ml-connectivity-total} shows the average results after $\epsilon=20$ with an validation loss of $\text{min}_{\epsilon}[\langle l_\text{c,val} \rangle]=0.165  \pm 0.001$ (corresponding to a maximal validation accuracy $\text{max}_{\epsilon}[\langle a_\text{c,val} \rangle]= 92.702\% \pm 0.001$).
At first glance, the figure seems to indicate a great success: from the $31000$ states present in $\tau$, $11510.6$ have been correctly classified as non-spanning (i.e., $N\rightarrow N'$), and $17205.9$ as spanning ($S\rightarrow S'$) while only $1223.1$ are wrongly labeled as non-spanning ($S\rightarrow N'$) and $1059.41$ as spanning ($N\rightarrow S'$) \footnote{We note that these numbers are not integers since they are computed as averages over the $10$ independent training runs as mentioned in section \ref{sec:supervised}}. Overall, we would conclude that $92.6\%$ of all test states are correctly classified while $7.4\%$ are wrong. 
However, from the full percolation analysis for $\mathcal{T}\cup\mathcal{V}$ shown in Fig.\ \ref{fig:percolation}, we know that there are $92.7\%$ of states without a spanning cluster below $p_c(L)$ while $7.3\%$ of states, equivalent to $876$ samples, already contain a spanning cluster. Similarly, for $p>p_c(L)$, $94.8\%$ of states, equivalent to $936$ samples, are spanning and $5.2\%$ are not. At $p_c(L)=0.585$, we furthermore have $518$ spanning and $482$ non-spanning states. Hence in total, we expect $2812$ wrongly classified states.
Since the last number is decisively close to the actual number of $2282.5$ of misclassified states, this suggests that it is precisely the spanning states below $p_c(L)$ and the non-spanning ones above $p_c(L)$ which the DL network is unable to recognize. 
Let us rephrase for clarity: it seems that the DL CNN, when trained in whether a cluster is spanning or non-spanning, completely disregards this information in its classification outputs. 

\subsection{\label{sec:sup-connectivity-threshold}Density-resolved study of spanning/non-spanning close to $p_c(L)$}

\begin{figure}[tb]
    (a)\includegraphics[width=0.9\columnwidth]{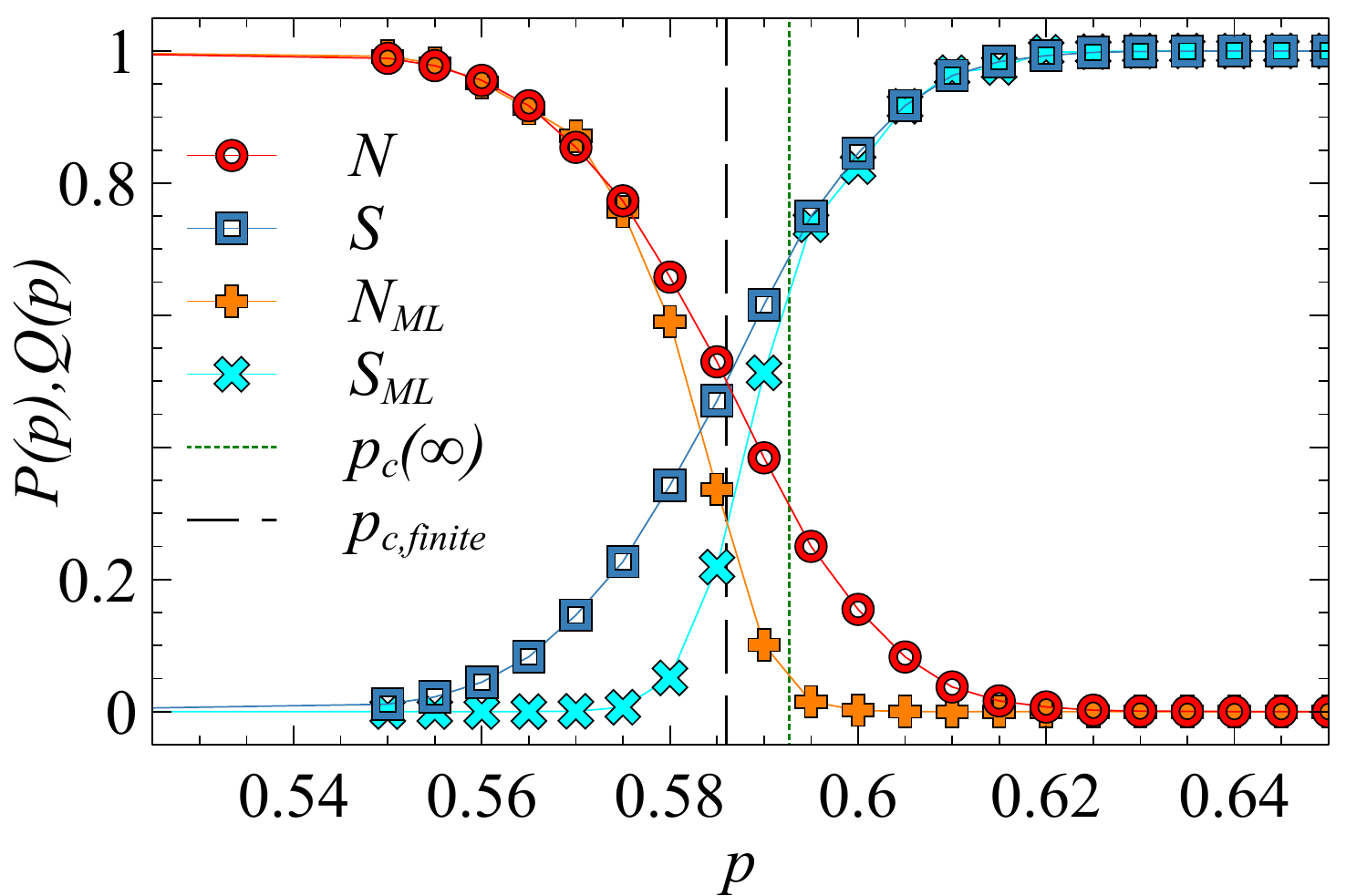}\\
    (b)\includegraphics[width=0.9\columnwidth]{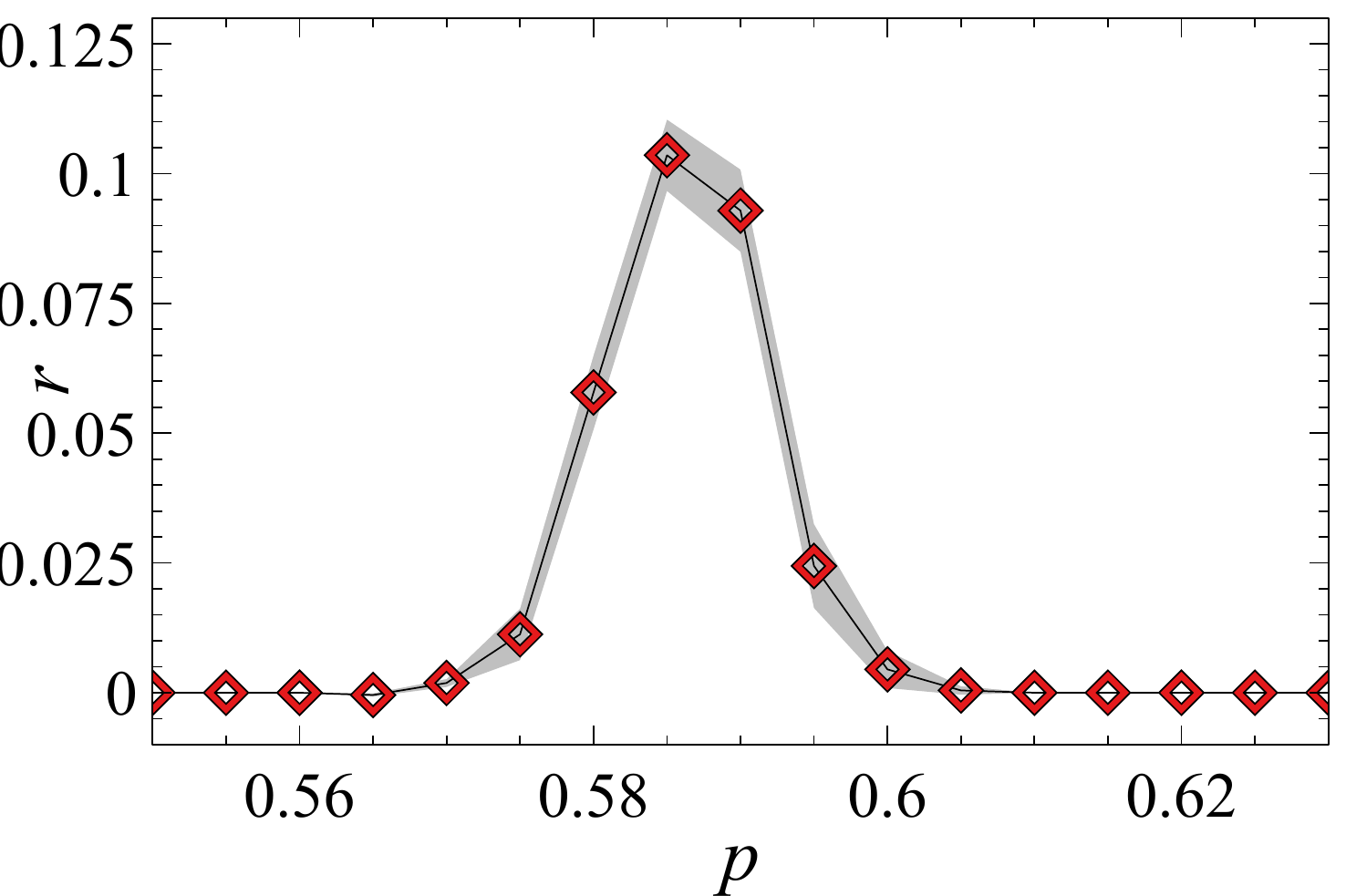}
     \caption{
     (a) The blue curve (red curve) shows the probability to have a spanning (non-spanning) sample in the training dataset. The cyan (orange) curve gives us the prediction of probability to have a spanning (non-spanning) sample, according to the trained network. 
     (b) Dependence of the Pearson correlation coefficient $r$ on the density $p$ for classification according to spanning/non-spanning. The confidence interval is indicated in gray.
     %
     In both (a) and (b), The lines connecting the symbols are only a guide to the eye.
    }
    \label{fig:ml-connectivity-p}
\end{figure}

In order to understand the behavior observed in the last section, we now reexamine the result of Fig.\ \ref{fig:ml-connectivity-total} by analyzing the ML-predicted probabilities, $P_\text{ML}(p)$. In Fig.\ \ref{fig:ml-connectivity-p}, we show both $P_\text{ML}(p)$ as well as $P(p)$; the latter having been obtained by the Hoshen-Kopelman algorithm, cf.\ Fig.\ \ref{fig:percolation}(a).
While the $P(p)$ and $P_\text{ML}(p)$ curves --- and of course also the corresponding $Q(p)$ and $Q_\text{ML}(p)$ --- appear qualitatively similar, they are nevertheless not identical and 
the slopes of $P_\text{ML}(p)$, $Q_\text{ML}(p)$ are different. We emphasize that the slopes are important for determining the universality class of a second-order phase transition via finite-size scaling \cite{Kirkpatrick1992AProblem}. 
Since we know for each image whether it percolates or not, we can also check how well the ML predictions worked by considering the covariance. Let $\zeta(\psi_i(p))=0$ when there is no percolating cluster in the state $\psi_i(p)$ while $\zeta(\psi_i(p))=1$ if there is. Similarly, we define $\zeta_\text{ML}(\psi_i(p))$ for the prediction by the DL network. Then $\text{cov}(\zeta,\zeta_\text{ML})(p)$ measures the covariance of states being found to span by percolation \emph{and} by ML for given $p$. In Fig.\ \ref{fig:ml-connectivity-p}(b) we show the normalized result, i.e., the Pearson coefficient $r_{\zeta,\zeta_\text{ML}}(p)=\text{cov}(\zeta,\zeta_\text{ML})(p)/ [ \sigma_{\zeta}(p) \sigma_{\zeta_\text{ML}}(p) ]$, where $\sigma_{\zeta}$ and $\sigma_{\zeta_\text{ML}}$ are the standard deviations of the percolation results and the ML predictions. We see that in the transition region,  $r_{\zeta,\zeta_\text{ML}} \lesssim 0.12$ which is very far from the maximally possible value $1$.
This suggest that while the ML predictions are not simply random, they are also not very much better than random.

\begin{figure*}[tb]
    \centering%
    (a) 
    \begin{minipage}{0.35\textwidth}\hspace*{3ex}{\centering $p < p_c$}\\[2ex]
        \hspace*{5ex} $p=0.56$ \hspace*{13ex} $p=0.57$ \\
        \includegraphics[width=0.49\columnwidth]{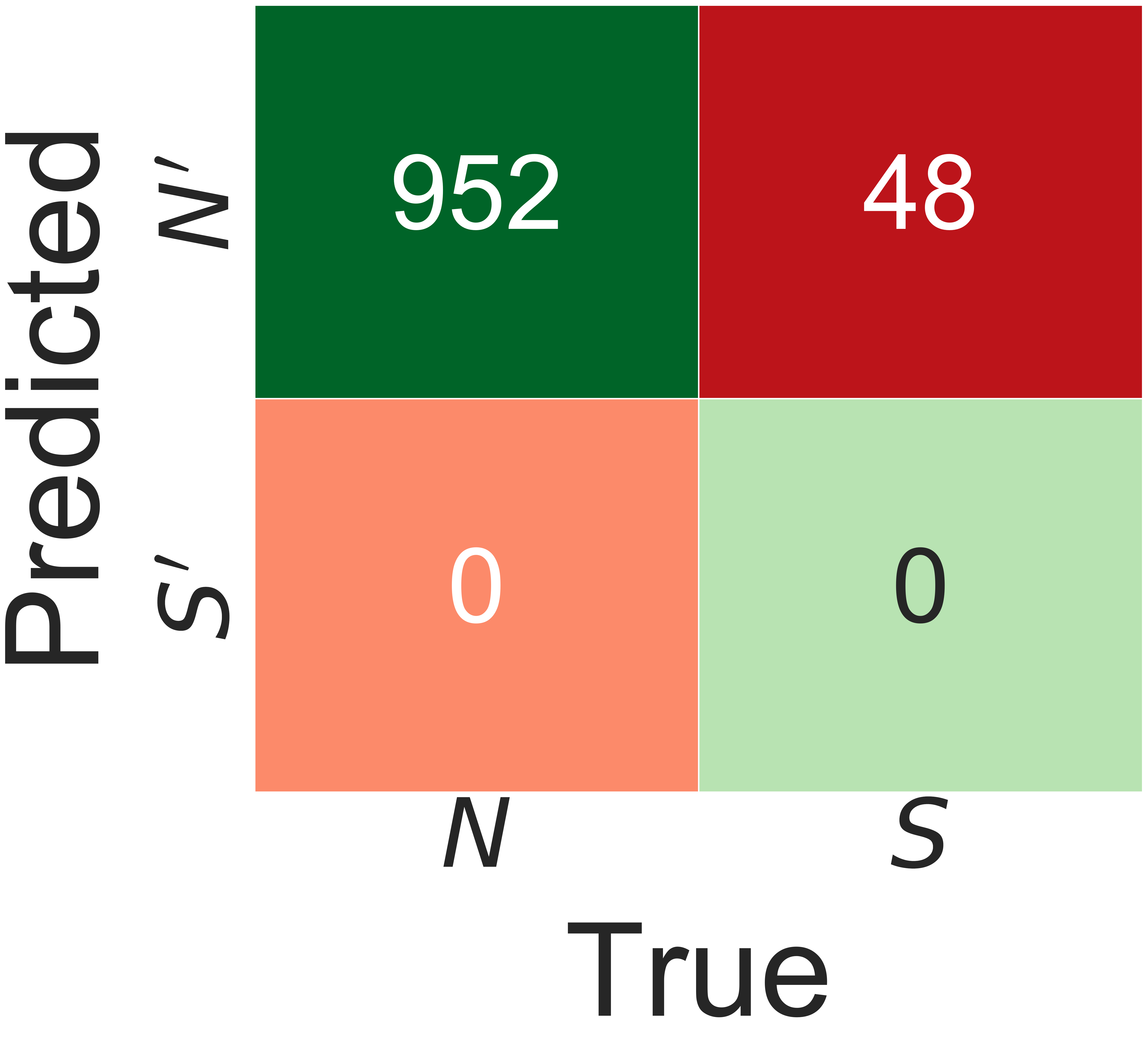}
        \includegraphics[width=0.49\columnwidth]{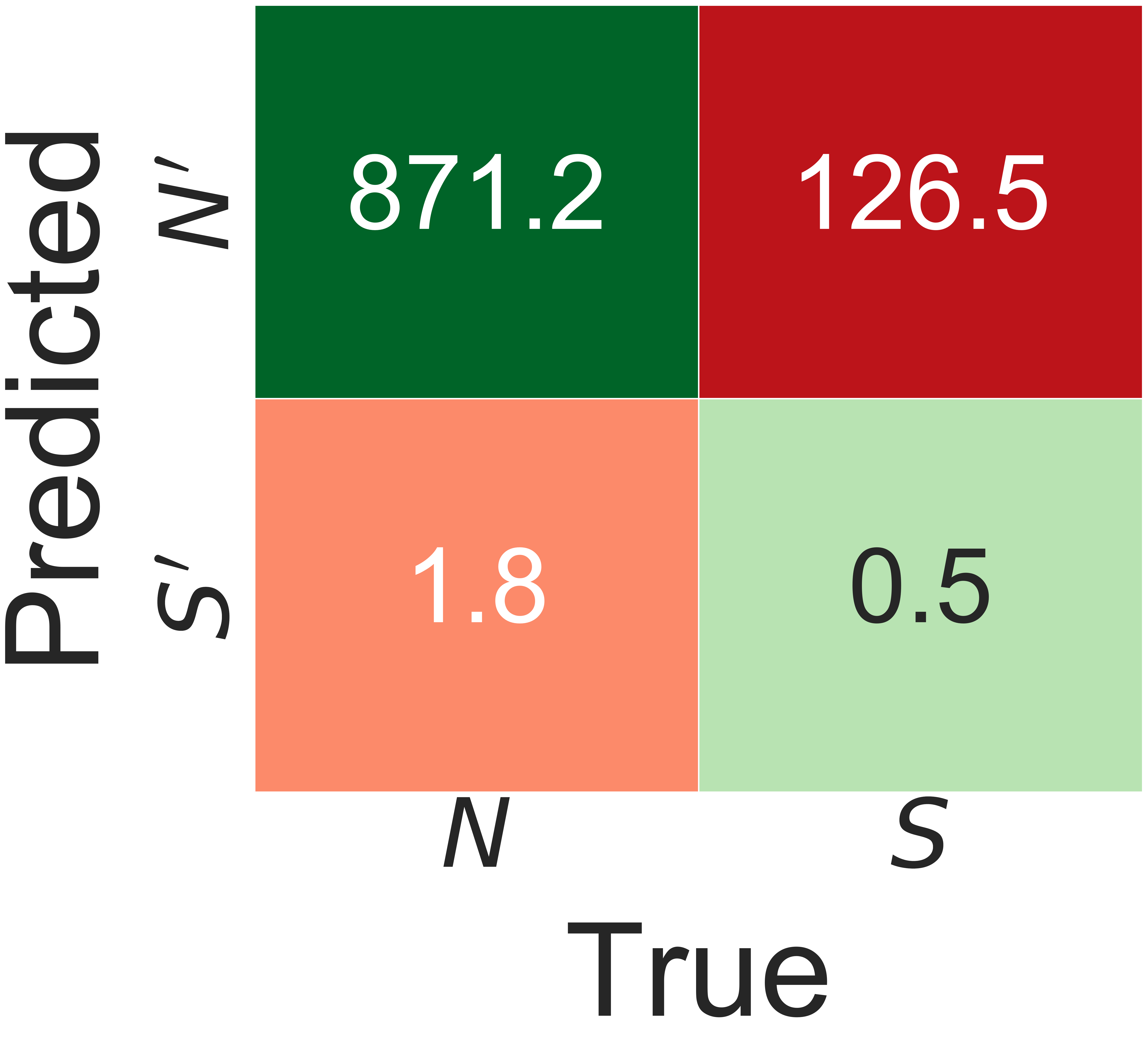}
        \hspace*{5ex} $p=0.565$ \hspace*{13ex} $p=0.575$ \\
        \includegraphics[width=0.49\columnwidth]{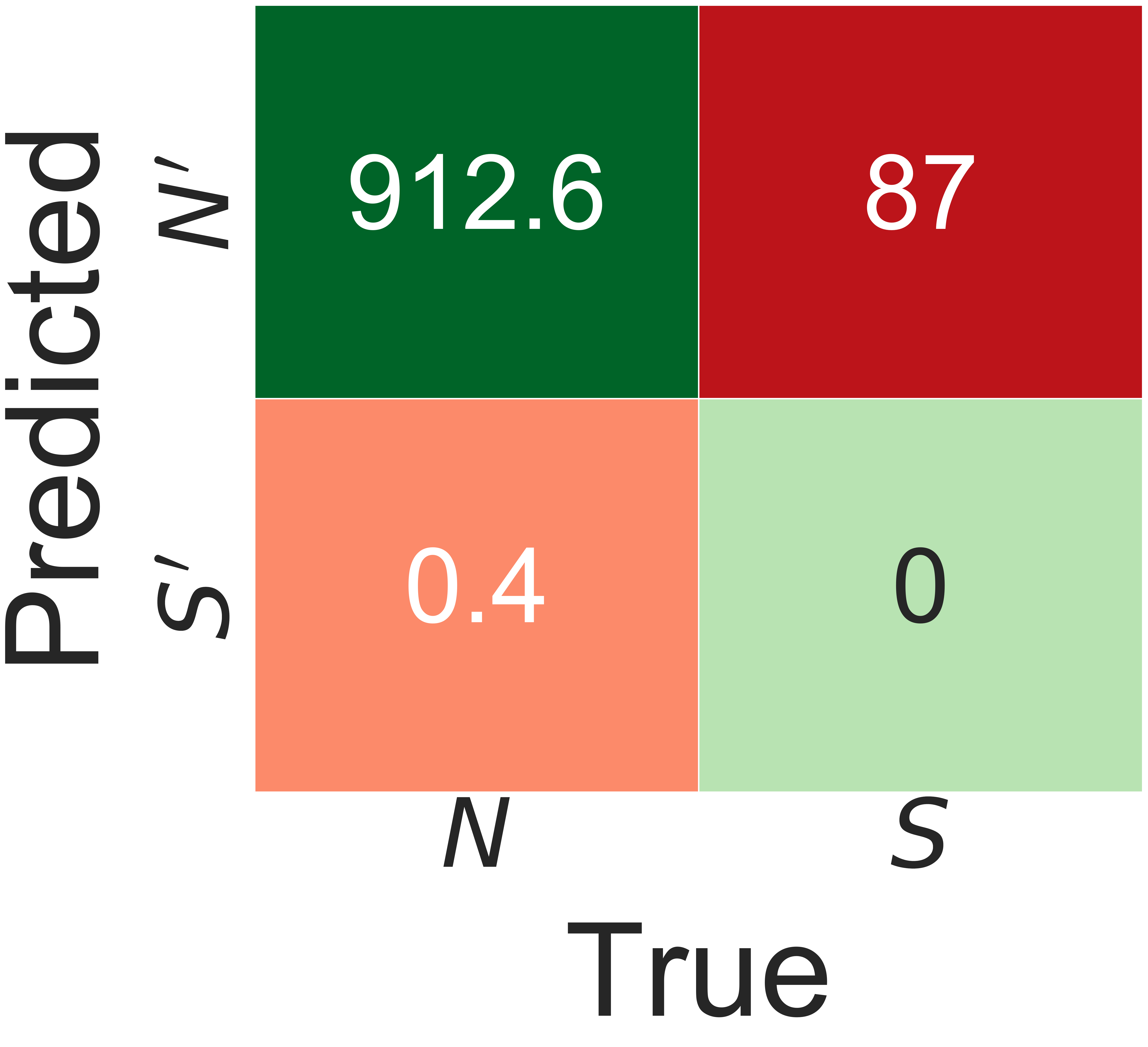} 
        \includegraphics[width=0.49\columnwidth]{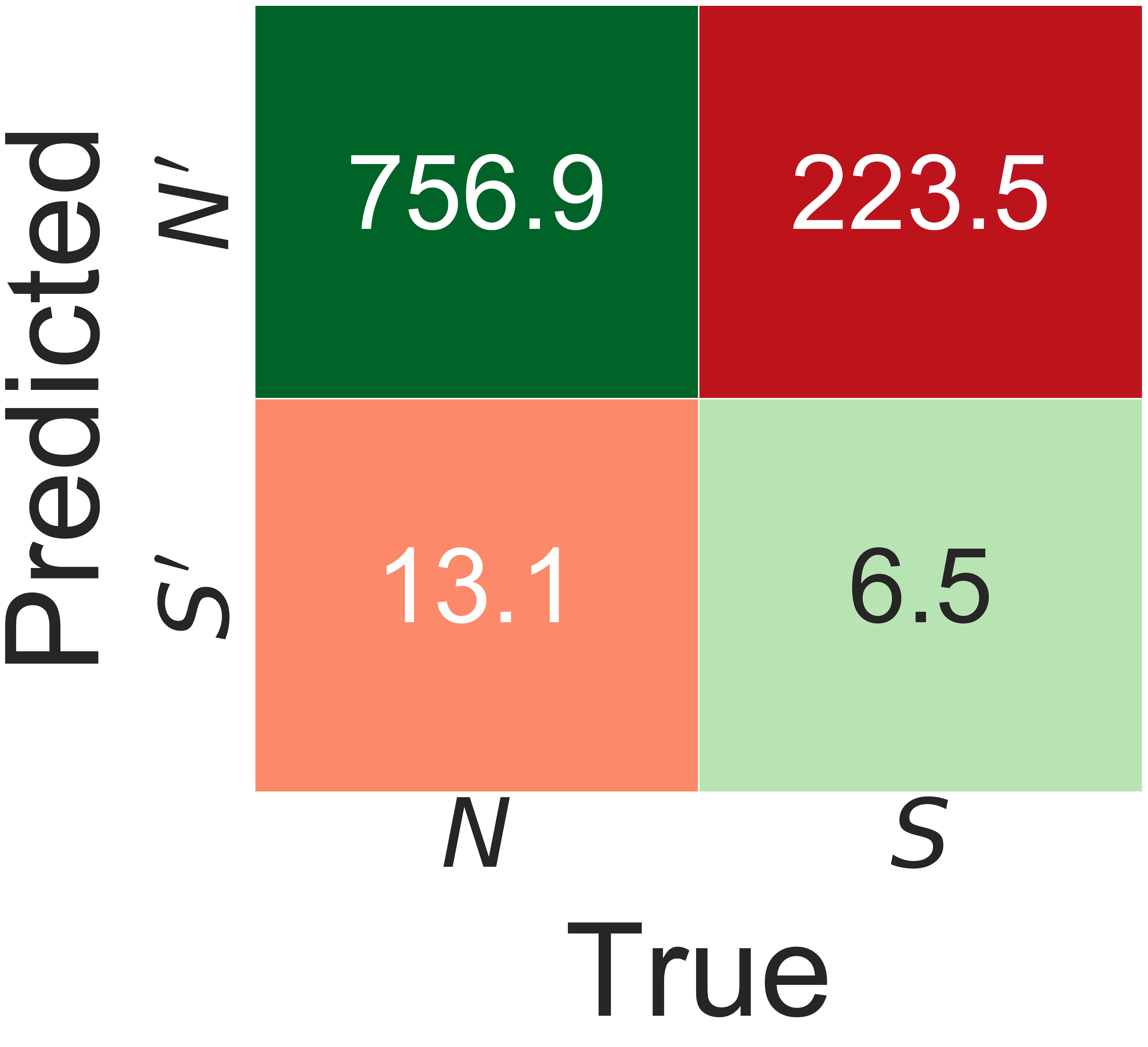}  
    \end{minipage}
    (b) 
    \begin{minipage}{0.19\textwidth}\hspace*{4ex}{\centering $p \sim p_c$}\\[2ex]
        \hspace*{5ex} $p=0.58$ \\
        \includegraphics[width=0.9\columnwidth]{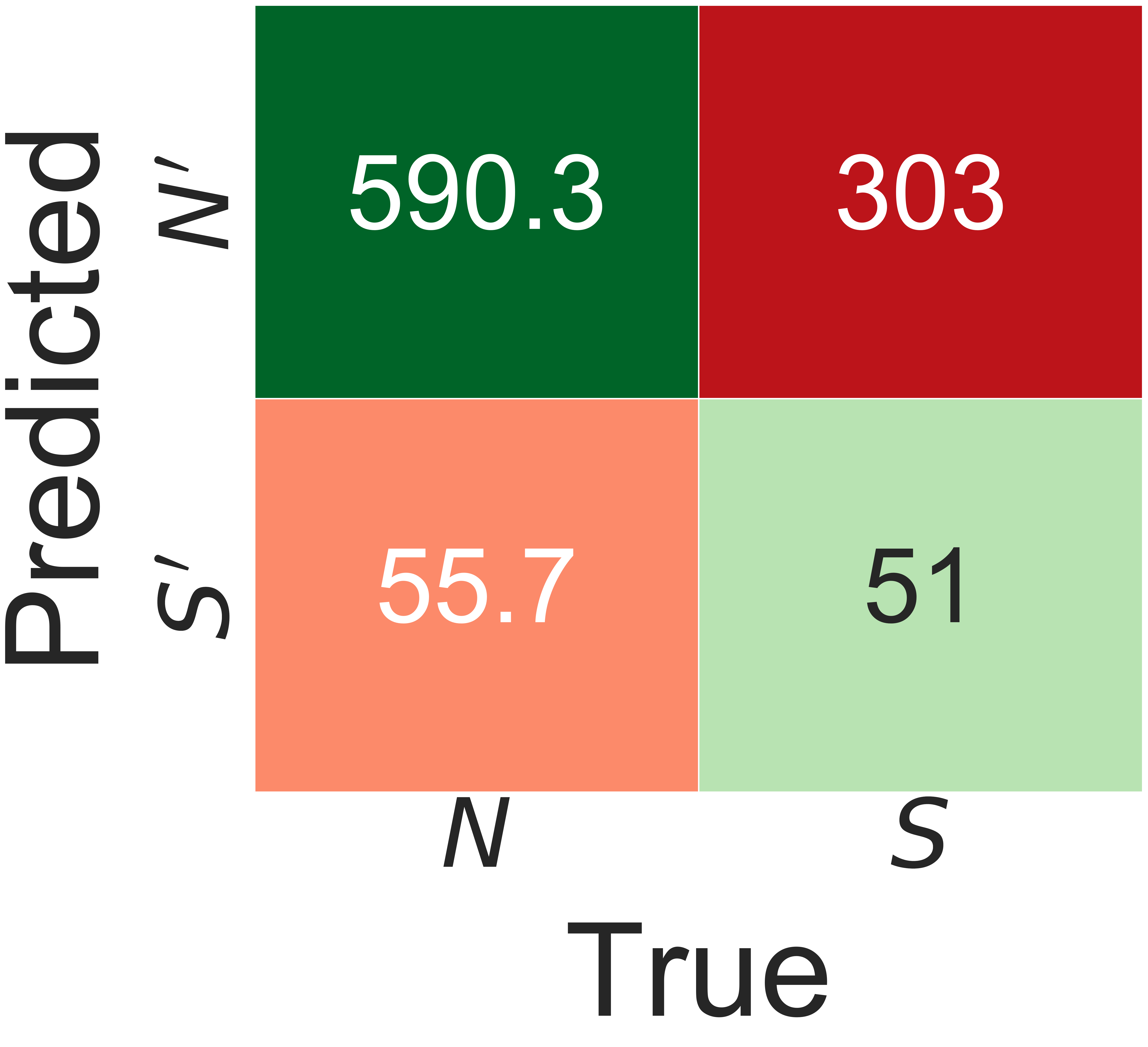}\\
         \hspace*{5ex} $p=0.585$ \\
        \includegraphics[width=0.9\columnwidth]{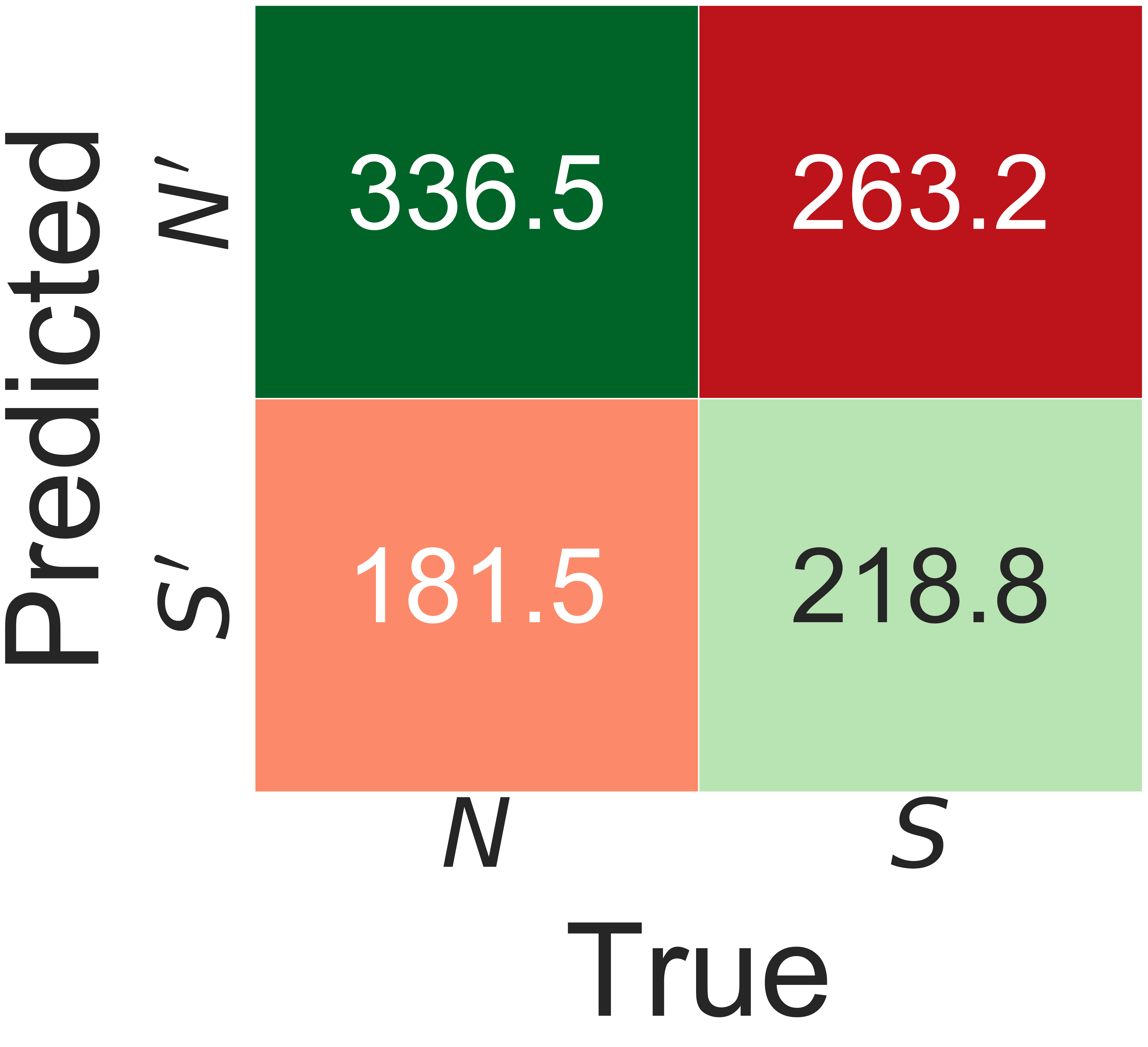}
    \end{minipage}
    (c)
    \begin{minipage}{0.35\textwidth}\hspace*{3ex}{\centering $p > p_c$}\\[2ex]
        \hspace*{5ex} $p=0.59$ \hspace*{13ex} $p=0.6$ \\
        \includegraphics[width=0.49\columnwidth]{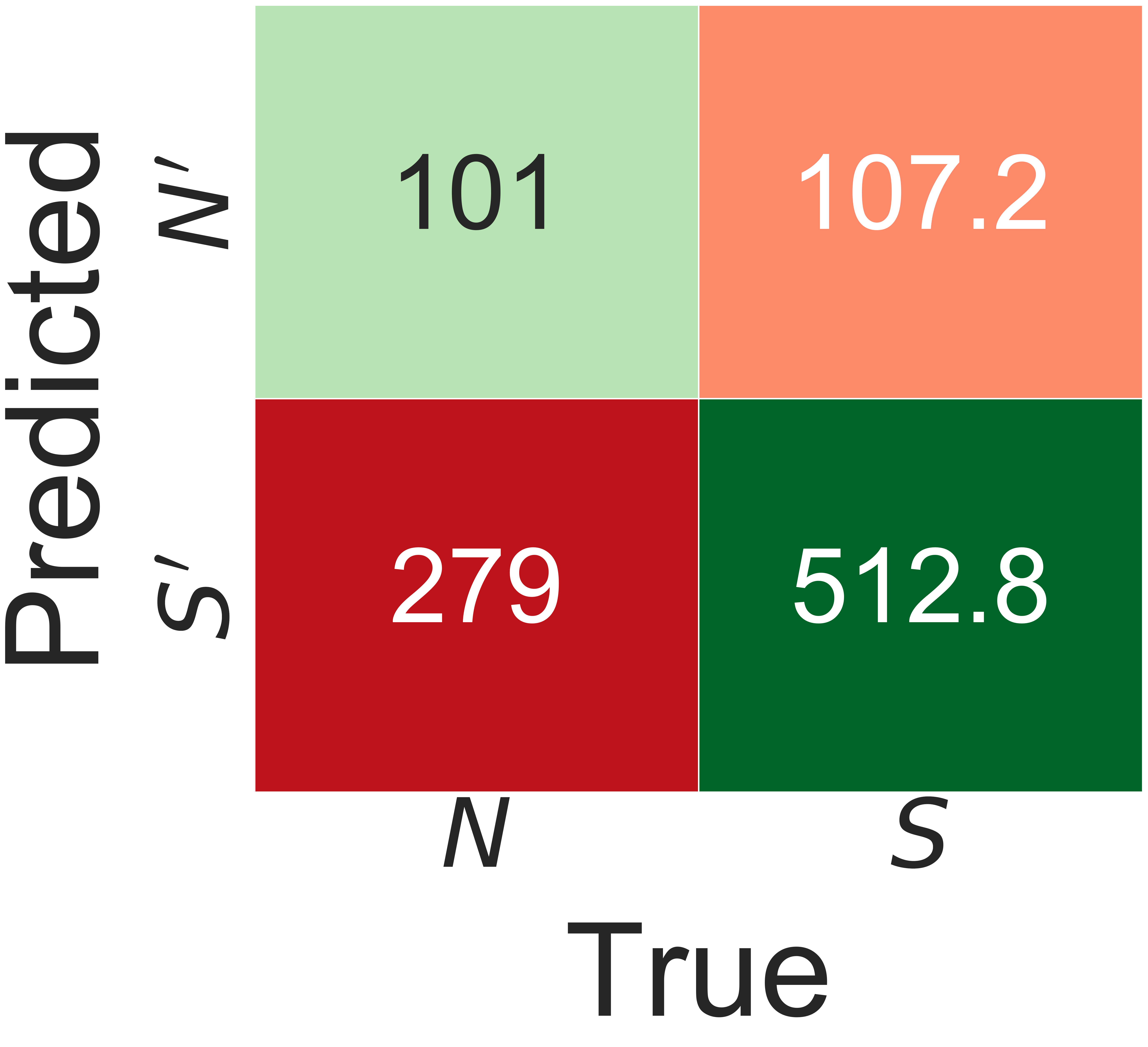}
        \includegraphics[width=0.49\columnwidth]{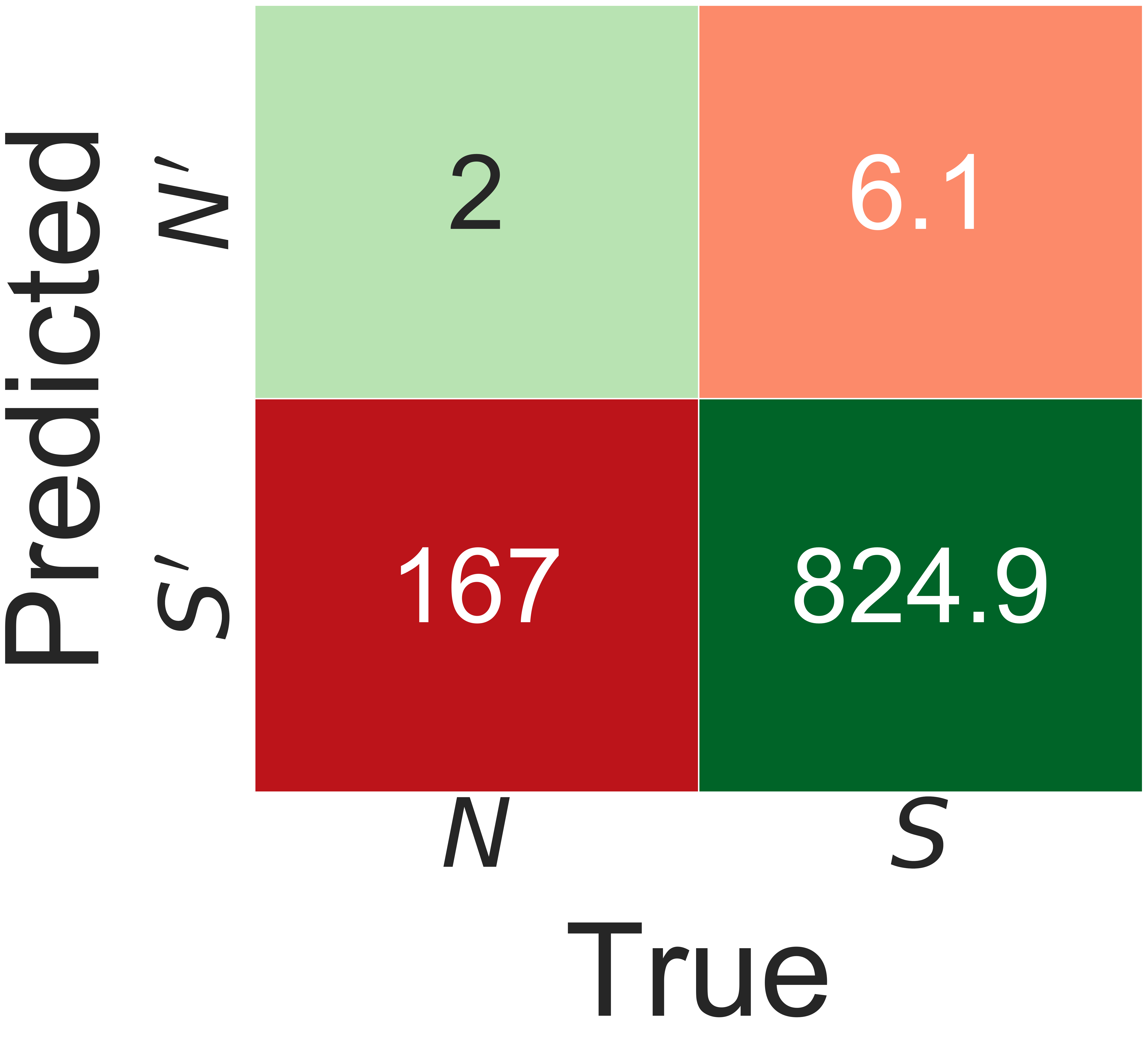}
        \hspace*{5ex} $p=0.595$ \hspace*{13ex} $p=0.605$ \\
         \includegraphics[width=0.49\columnwidth]{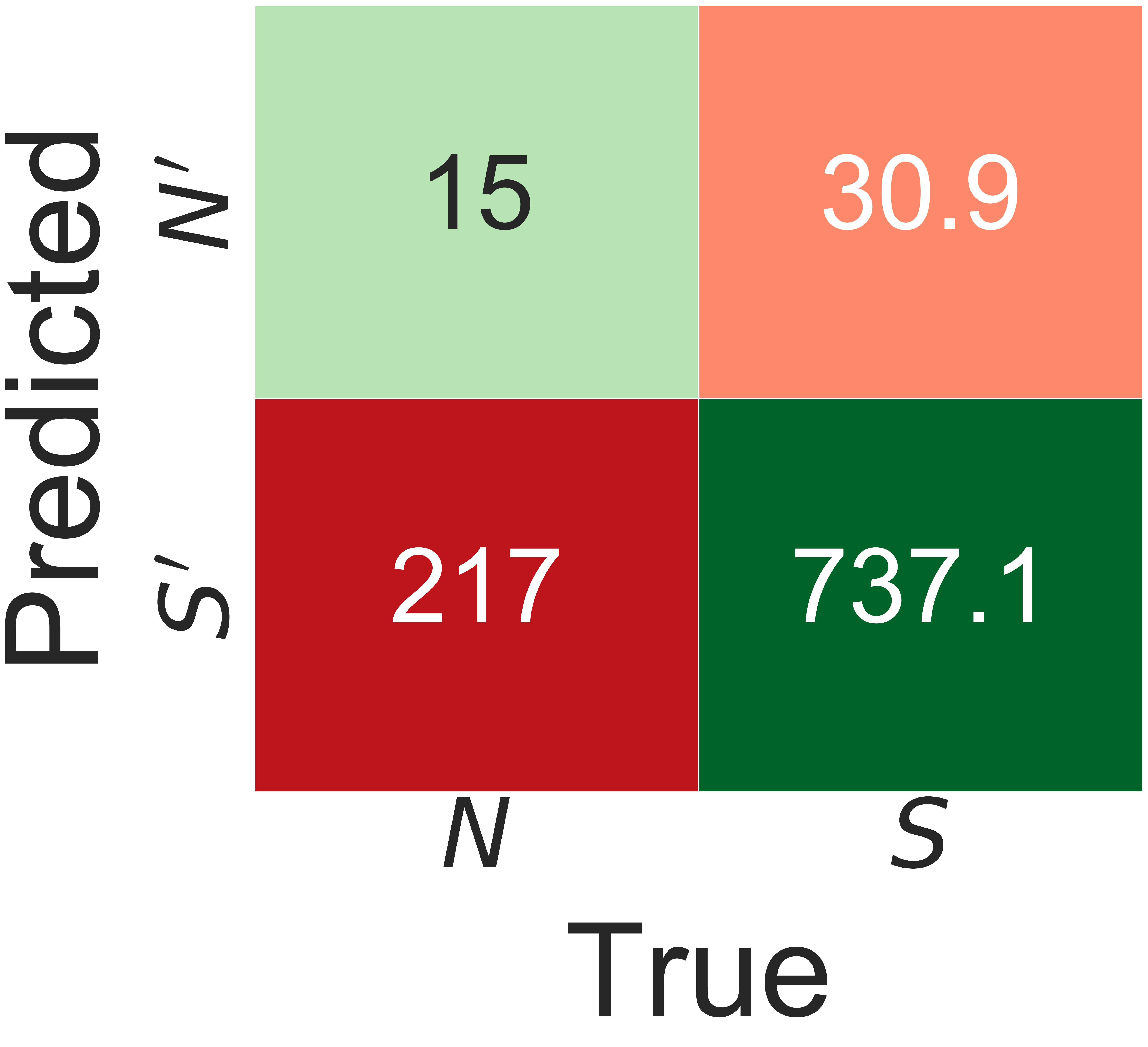}
        \includegraphics[width=0.49\columnwidth]{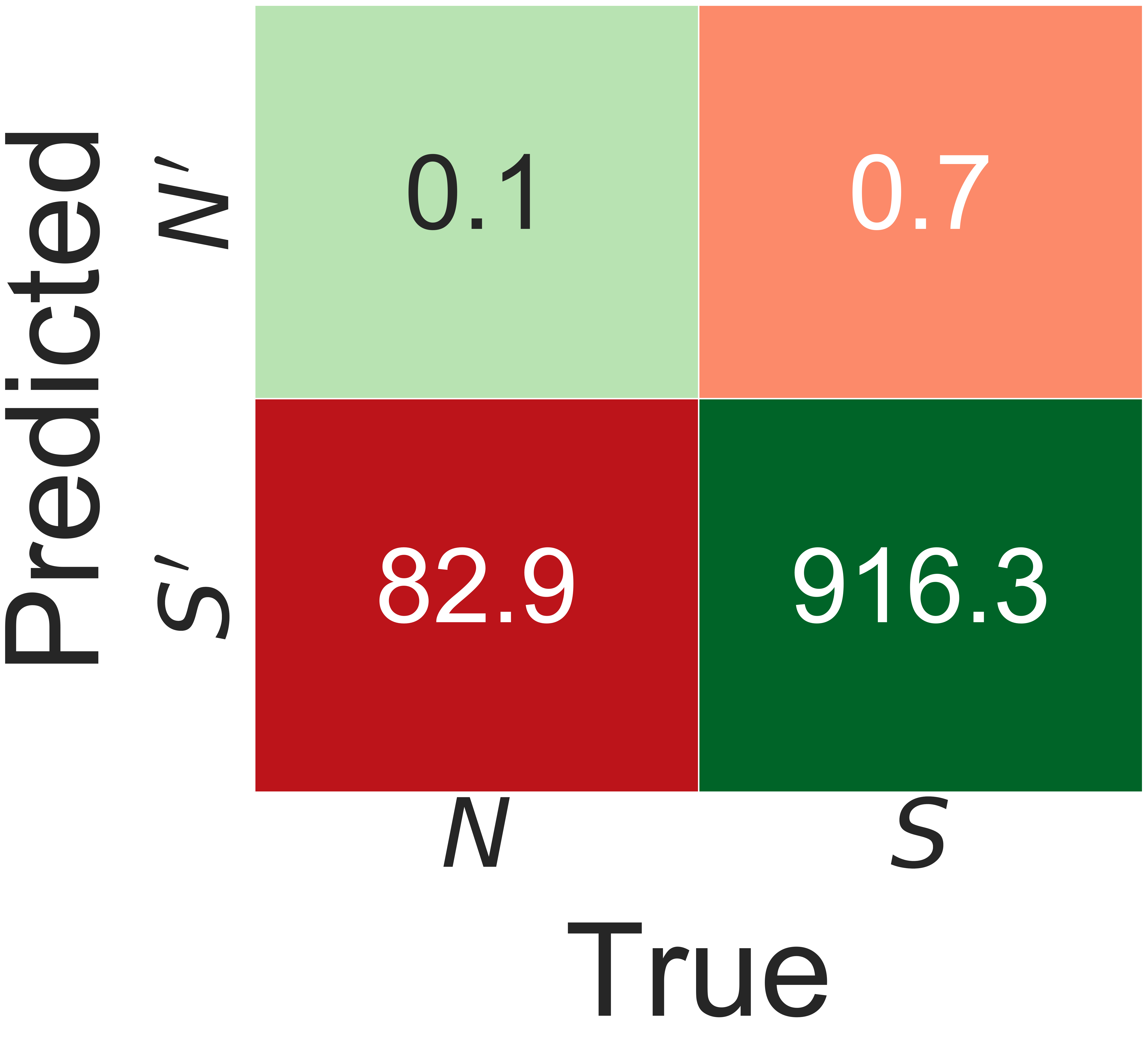}
    \end{minipage}
    \caption{Confusion matrices showing the predictions of the trained network Fig.\ \ref{fig:ml-connectivity-total} in a region $p=[0.56,0.605]$  comprising $p_c$, with (a) for predictions made before the percolation threshold, (b) in the threshold region and (c) after the percolation threshold. Each confusion matrix is an average of the predictions made by the  $10$ trained models shown in Fig.\ \ref{fig:ml-connectivity-total}.}
    %
    \label{fig:ml-connectivity-cm}
\end{figure*}

Let us now study the classification into spanning/non-spanning states in detail for each $p$. Figure \ref{fig:ml-connectivity-cm} and Table \ref{tab:ml-connectivity-table} show a comparison of the classification for the ten $p$ values $0.56$ to $0.605$. 
\begin{table*}[tb]
\centering
 \begin{tabular}{
      l
      S[table-format=-3.2]
      S[table-format=-3.2]
      S[table-format=-3.2]
      S[table-format=-3.2]
      S[table-format=-3.2]
      S[table-format=-3.2]
      S[table-format=-3.2]
      S[table-format=-3.2]
      S[table-format=-3.2]
      S[table-format=-3.2]
      S[table-format=-3.2]
      S[table-format=-3.2]
      S[table-format=-3.2]}  
 \hline \hline
 & &  \multicolumn{2}{c}{$N$} & \multicolumn{2}{c}{$S$} & \multicolumn{2}{c}{$(S\rightarrow S')$} & \multicolumn{2}{c}{$(S\rightarrow N')$} & \multicolumn{2}{c}{$(N \rightarrow S')$} & \multicolumn{2}{c}{$(N\rightarrow N')$} \\
  $p$ & $n$ &  \# & \%  &  \#  & \%  & $\langle \# \rangle$ & $ \langle \% \rangle$ &$\langle \# \rangle$ &$ \langle \% \rangle$& $\langle \# \rangle$ & $ \langle \% \rangle$ & $\langle \# \rangle$ & $ \langle \% \rangle$\\ 
 \hline
 0.56  & 1000 & 952 & 95.2 & 48  &  4.8 &   0.0  &   0.0  &  48.0 &  4.8  &   0.0 &  0.0  & 952.0 & 95.2\\ 
 0.565 & 1000 & 913 & 91.3 & 87  &  8.7 &   0.0  &   0.0  &  87.0 &  8.7  &   0.4 &  0.0  & 912.6 & 91.3\\
 0.57  & 1000 & 873 & 87.3 & 127 & 12.7 &   0.5  &  0.1  & 126.5 & 12.7  &   1.8 &  0.2  & 871.2 & 87.1\\
 0.575 & 1000 & 770 & 77.0 & 230 & 23.0 &   6.5  &  0.7  & 223.5 & 22.4  &   13.1 & 1.3  & 756.9 & 75.7\\
 0.58  & 1000 & 646 & 64.6 & 354 & 35.4 &  51.0  &  5.1  & 303.0 & 30.3  &   55.7 &  5.6 & 590.3 & 59.0\\ 
 0.585 & 1000 & 518 & 51.8 & 482 & 48.2 & 218.8  &  21.9 & 263.2 & 26.3  &  181.5 & 18.2 & 336.5 & 33.6\\ 
 0.59  & 1000 & 380 & 38.0 & 620 & 62.0 & 512.8  &  51.3 & 107.2 & 10.7  &  279.0 & 27.9 & 101.0 & 10.1\\ 
 0.595 & 1000 & 232 & 23.2 & 768 & 76.8 & 737.1  &  73.7 &  30.9 &  3.1  &  217.0 & 21.7 &  15.0 & 1.5\\ 
 0.60  & 1000 & 169 & 16.9 & 831 & 83.1 & 824.9  &  82.5 &   6.1 &  0.6  &  167.0 & 16.7 &   2.0 & 0.2\\ 
 0.605 & 1000 &  83 & 8.3  & 917 & 91.7 & 916.3  &  91.6 &   0.7 &  0.1  &  82.9 &  8.3  &  0.1  & 0.0\\ 
 \hline \hline
 \end{tabular}
 \caption{ \label{tab:ml-connectivity-table}
Predictions of the trained network on the test data set $\tau$ for $p \in [0.56,0.605]$. $N$ and $S$ denote the respectively the number of non-spanning and the number of spanning samples in $\phi$. The four following columns $(S\rightarrow S')$, $(S\rightarrow N')$, $(N\rightarrow S')$, and $(N\rightarrow N')$ gives the averaging of 10 independent prediction runs.}
 \end{table*}
We see, e.g., that for $p=0.56, 0.565, 0.57, 0.575 < p_c(L)\sim 0.58, 0.585$, $485$ of $487$ samples, which are already spanning, have been misclassified as non-spanning. Similarly, for $p= 0.59, 0.595, 0.6, 0.605 > p_c(L)$, $7545.9$ of in total $864$ still non-spanning samples are classified as spanning. These results are similar whether one considers a typical sample or the averaged result. Hence, contrary to the supposed success of Fig.\ \ref{fig:ml-connectivity-total}, we now find that the seemingly few misclassified states of Fig.\ \ref{fig:ml-connectivity-total} are indeed precisely those which represent the correct physics. Saying it differently, the ML process seems to have led to a DL network which largely disregards the characteristic of spanning clusters and just uses the overall density of occupied vs.\ non-occupied sites to ascertain the phases. Of course, this is the wrong physics when considering percolation.

\subsection{\label{sec:sup-tests}Testing the accuracy of the DL network}

\begin{figure*}[tb]
    \centering%
    (a)\hspace*{-2ex}
    \begin{minipage}{0.33\textwidth}
    \raisebox{5ex}{\includegraphics[width=0.33\columnwidth]{Spanning_non_spanning_test/pc_L100_line_0.5_3lines_col.pdf}}
    \includegraphics[width=0.57\columnwidth]{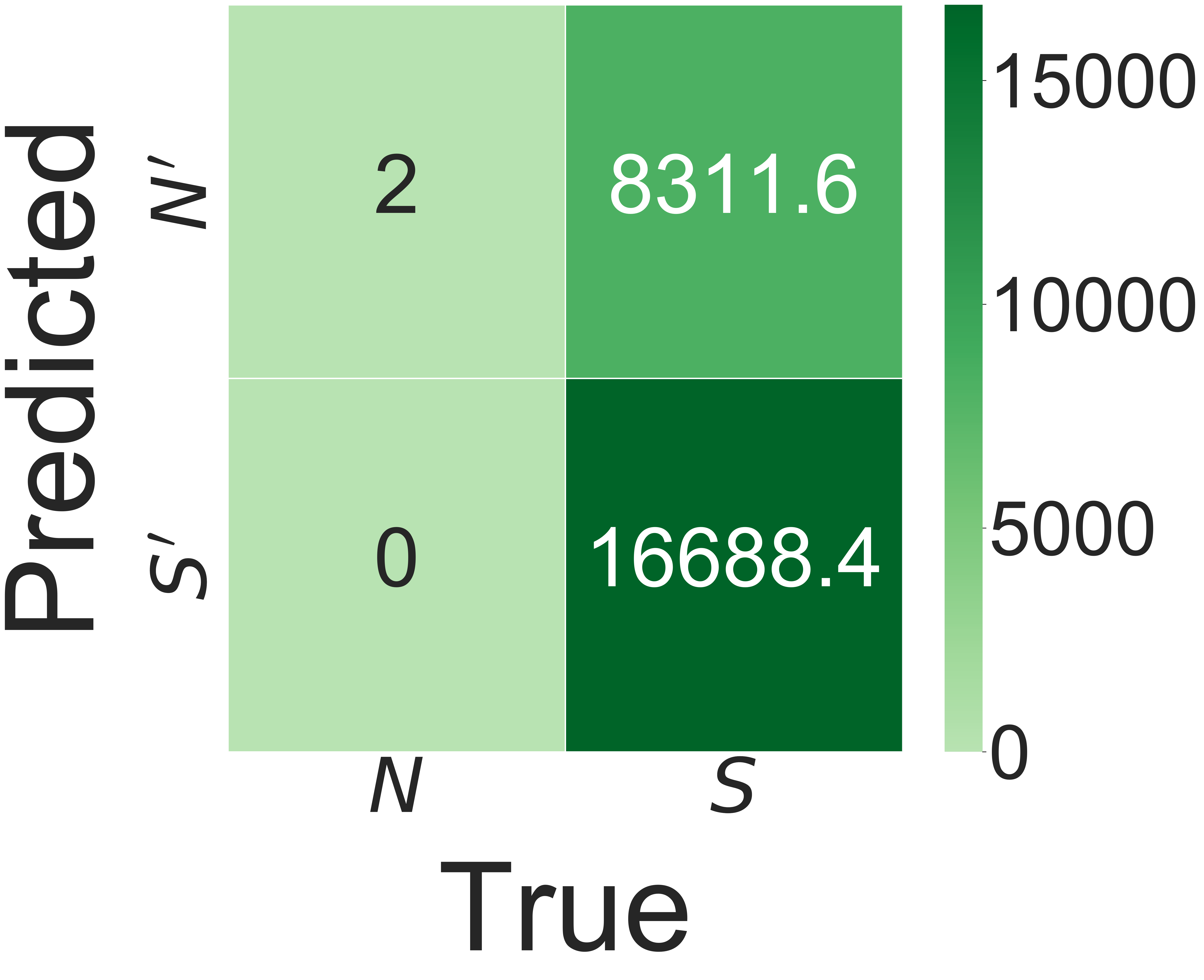}
    \end{minipage}\hspace*{-1ex}
    (b)\hspace*{-2ex}
    \begin{minipage}{0.33\textwidth}
    \raisebox{5ex}{\includegraphics[width=0.33\columnwidth]{Spanning_non_spanning_test/pc_L100_randpath_0.5_3lines_col.pdf}}
    \includegraphics[width=0.57\columnwidth]{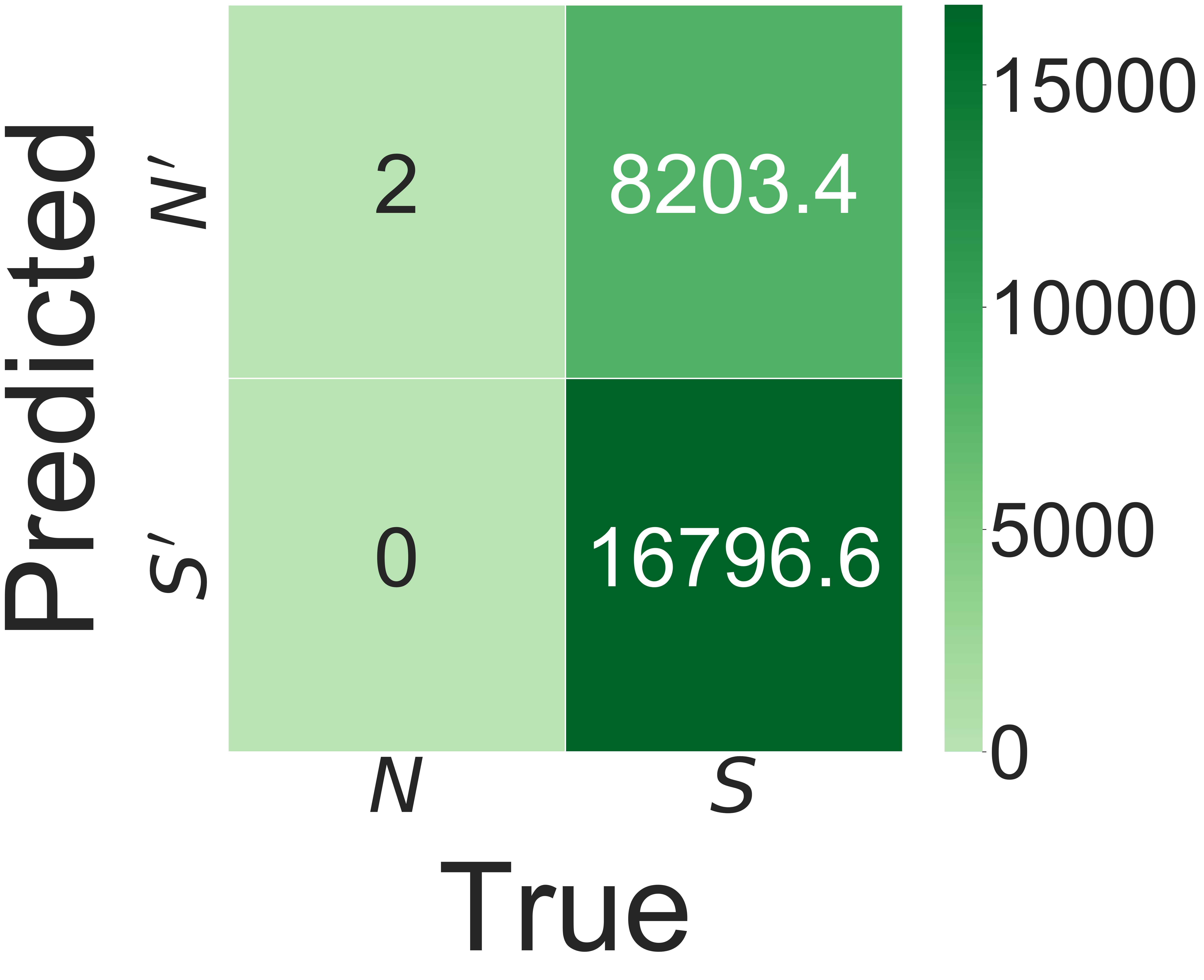}
    \end{minipage}\hspace*{-1ex}
    (c)\hspace*{-2ex}
    \begin{minipage}{0.33\textwidth}
    \raisebox{5ex}{\includegraphics[width=0.33\columnwidth]{Spanning_non_spanning_test/anticross_pc_0_0_0_0_0__p0.5_L100_s15416711_nc592_smc265_n592_bw.png}}
    \includegraphics[width=0.57\columnwidth]{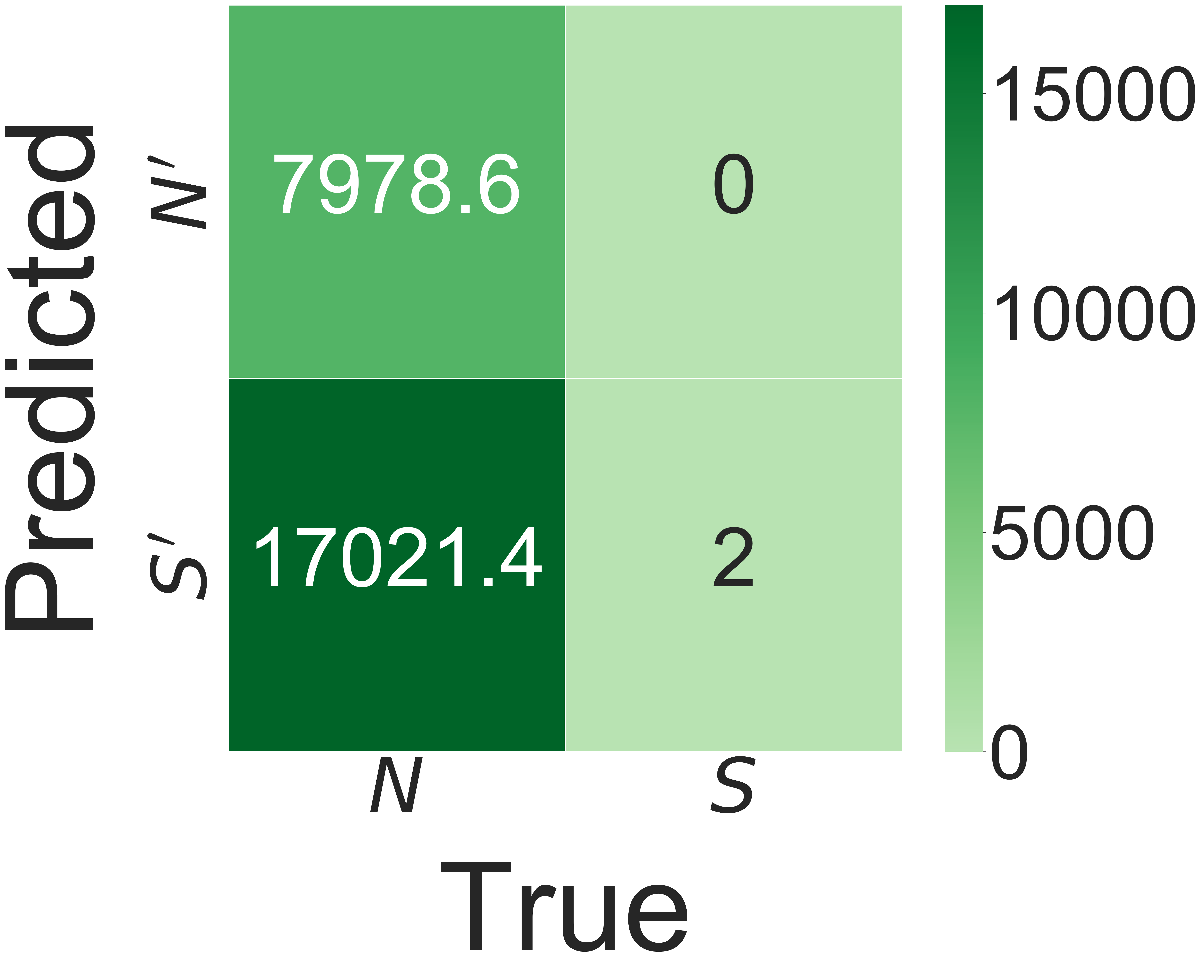}
    \end{minipage}
    \caption{Sample states for the three special test sets (a) $\tau_\text{sl}$ with added straight spanning lines, (b) $\tau_\text{rw}$ with spanning random walks and (c) $\tau_\text{fb}$ with the non-spanning firebreaks. In each case, the right plots gives the confusions matrices obtained from the DL model previously trained in a spannin vs.\ non-spanning classification. In all cases, the density is strictly $p=0.5$.
    }
    \label{fig:ml-class-test-images}
\end{figure*}

The difficulties that the trained DL network has with recognizing whether a state contains a percolating cluster or not can be made more explicit. In section \ref{sec:test-data}, we had generated three test sets $\tau_S$ for this purpose. Namely, percolating states even for $p< p_c(L)$ by adding (i) a straight line and (ii) a random walk of connecting sites as well as  for $p>p_c(L)$ (iii) the firebreak states of percolation-prohibiting random unoccupied sites. We now use these sets and feed them independently as test sets to the DL network. 
Figure \ref{fig:ml-class-test-images} shows the three confusion matrices obtained when classifying for spanning vs.\ non-spanning. In Fig.\ \ref{fig:ml-class-test-images}(a+b), we see that the network completely misclassifies the spanning datasets $\tau_\text{Sl}$ and $\tau_\text{rw}$. The two correctly identified non-spanning images are just the two such states added to each of the data sets to show that the network is still performing.
Similarly, in Fig.\ \ref{fig:ml-class-test-images}(c), we see that this time the network cannot correctly identify the non-spanning samples in $\tau_\text{fb}$. Again, the two samples correctly identified are the ones without the firebreak. 

\section{\label{sec:conclusions}Conclusions}

Let us briefly summarize what we have achieved thus far. 
We showed that when looking at $p$, classification and regression techniques for percolation states allow us to obtain good recognition with near-perfect $\langle a_{c,val} \rangle = 99.323\% \pm 0.003$) for classification and near-zero $\langle l_{r,val} \rangle =0.000062 \pm 0.000012$ average mean-square loss for regression. Confusion matrices are heavily diagonal for classification while prediction curves for regression are similarly convincing. These results are in good agreement with previous similar such studies \cite{Zhang2019,Shen2021SupervisedPercolation,Cheng2021MachineModel}, which should not come as a surprise: the information about $p$ is of course directly enclosed in each state. We emphasize that our approach is already somewhat more challenging than these previous works since instead of just asking the DL network to identify the two phases $p < p_c$ and $p> p_c$, we also successfully identify all $31$ distinct densities $p$.
Using $\langle\xi(p)\rangle$ instead of $p$ to identify the $31$ densities also works quite well, but is again expected since $\langle\xi(p)\rangle$ merely acts as a new set of supervised labels.

Problems emerge when we use the computed correlation length $\xi$ for each state and try classification and regression with these correlation lengths. Having thus explicitly removed any connection with the $p$ values, we nevertheless find that the resulting confusion matrices, fidelity curves and losses are all of much less quality than before. Instead, it seems that the density $p$ information is still the overriding measure used by the DL network to arrive at its outputs (cf.\ Figs.\ \ref{fig:ml-correlation-classification-pdf} and \ref{fig:ml-correlation-regression}) \footnote{We emphasize that we have spent considerable effort at making sure that this result is not due to erroneous information leakage \cite{Dawid2022ModernWittek}.}.
Rather, as we show in sections \ref{sec:sup-connectivity-total}--G, the DL network completely ignores whether a cluster is spanning or non-spanning, essentially missing the underlying physics of the percolation problem --- it seems to still use $p$ as its main ordering measure.

We believe that the root cause of the failure to identify the spanning clusters, or their absence, lies in the fundamentally local nature of the CNN: the filter/kernels employed in the {\sc ResNet}s span a few \emph{local} sites only \footnote{In addition to the {\sc ResNet18} used in the main text, we have also checked that {\sc ResNet34} yields a similar outcome.}. Hence it is not entirely surprising that such a CNN cannot correctly identify the essentially \emph{global} nature of spanning clusters. But it is of course exactly this global percolation that leads to the phase transition.

The reader might wonder why previous CNN studies of phases in other models, such as, e.g., the Ising-type models \cite{Carrasquilla2017}, the three-dimensional Anderson model and its topological variants \cite{doi:10.7566/JPSJ.89.022001}, have failed to find similar such issues. We think that this is because in the Ising case, the majority rule for spin alignment is not concerned with any globally spanning domain \cite{McCoy1973TheModel}, while in the Anderson-type models, it is the (typical) local density of states which can serve as order parameter \cite{Dobrosavljevic2003TypicalEffects}. In short, in these models a local property is indeed sufficient to distinguish their phases.

Of course ML aficionados might now want to suggest that extensions of local kernels are possible in CNNs. Indeed, one might, e.g., want to use CNNs in which large dilution parameters are employed to effectively make filters/kernels of manageable size while still spanning across a sizeable portion of the $L \times L$ size of each percolation state \cite{Al-Shabi2019Gated-DilatedScans}. But while this might solve the issue for fixed $L$ in the percolation case --- we have not tested it --- it does imply knowing that a global property is important to start with. This would rather diminish the relevance of DL as a tool for unbiased discovery in physics.

\begin{acknowledgments}
We thank B.\ \c{C}ivitcio\u{g}lu, M.\ Hilke, and T.\ Ohtsuki for discussions. 
D.B.\ is grateful for co-tutelle funding via the EUtopia Alliance. 
We gratefully acknowledge the University of Warwick Research Technology Platform (RTP Scientific Computing) and the Sulis Tier 2 HPC platform hosted by the RTP. Sulis is funded by EPSRC Grant EP/T022108/1 and the HPC Midlands+ consortium. 

\end{acknowledgments}
\bibliography{ML-DB.bib}

\end{document}